\documentclass[aps,10pt,nolongbibliography,prc,tightenlines,twocolumn,twoside,showpacs,superscriptaddress]{revtex4-1}
\usepackage[caption=false]{subfig}

\usepackage[hidelinks]{hyperref}
\usepackage{color}
\usepackage{graphicx}
\usepackage{siunitx}
\usepackage{amsmath,amssymb,amsfonts}
\usepackage{mathbbol}
\DeclareSymbolFontAlphabet{\mathbbmm}{bbold}
\DeclareMathAlphabet{\mathbbmsl}{U}{bbm}{m}{sl}
\DeclareSymbolFontAlphabet{\mathbb}{AMSb}%
\usepackage{bm}
\usepackage{bbm}
\usepackage{hyperref}
\usepackage{lineno}
\usepackage{slashed}
\usepackage{tikz}

\usepackage[normalem]{ulem}  

\def \be {\begin{equation} }
	\def \ee {\end{equation}}
\def \bes {\begin{subequations} }
	\def \ees {\end{subequations}}

\def \no {\nonumber}
\def \a {\alpha}
\def \b {\beta}

\def \d {\delta}
\def \e {\epsilon}
\def \g {\gamma}

\def \o {\omega}

\def \l {\lambda}
\def \m {\mu}
\def \n {\nu}

\def \pd {\partial}

\def \<{\langle}
\def \>{\rangle}
\def \+{\dagger}

\def \[{\left[}
\def \]{\right]}

\def \vp {\bm{p}}
\def \vq {\bm{q}}

\def \vk {\bm{k}}

\def \ve {\varepsilon}
\def \hn {\hat{n}}

\def \D {\Delta}
\def \G {\Gamma}

\def \<{\langle}
\def \>{\rangle}
\def \+{\dagger}

\def \[{\left[}
\def \]{\right]}

\def \sT {{\cal T}}
\def \srho 
{{\tilde \rho}}

\renewcommand{\sout}{\bgroup \color{red} \ULdepth=-.5ex \ULset}

\newcommand{\cleftsemicirc}{\put(3.5,2.5){\oval(8,8)[l]}\put(3.5,-1){\line(0,1){8}}\phantom{\circ}}
\newcommand{\crightsemicirc}{\put(1.5,2.5){\oval(8,8)[r]}\put(1.5,-1){\line(0,1){8}}\phantom{\circ}}

\begin{document}
	\title{Spectral properties and spin alignment of $\phi$ meson in QCD Matter}
		\author{Zhong-Yuan Sun
	}
	\affiliation{School of Physics and Electronics, Hunan University, Changsha 410082, China} 
	\affiliation{Hunan Provincial Key Laboratory of High-Energy Scale Physics
		and Applications, Hunan University, Changsha 410082, China}
	\author{You-Yu Li
	}
	\email{lyouyu2023@lzu.edu.cn}
	\affiliation{School of Physics and Electronics, Hunan University, Changsha 410082, China} 
	\affiliation{Hunan Provincial Key Laboratory of High-Energy Scale Physics
		and Applications, Hunan University, Changsha 410082, China}
        \affiliation{School of Physical Science and Technology, Lanzhou University, Lanzhou 730000, China}
	
	\author{Shuai Y.\,F.~Liu
	}
	\email{lshphy@hnu.edu.cn}
	\affiliation{School of Physics and Electronics, Hunan University, Changsha 410082, China} 
	\affiliation{Hunan Provincial Key Laboratory of High-Energy Scale Physics
		and Applications, Hunan University, Changsha 410082, China}


	\date{\today}
	
\begin{abstract}
As the spin alignment of a vector meson is predicted to be correlated with its spectral properties, we study  the spectral properties of the $\phi$ meson using two microscopic Lagrangians based either on chiral effective field theory or the quark-meson model. We calculate the self-energies and spectral functions of the $\phi$ meson for these two Lagrangians at the one-loop level within the Matsubara formalism of finite-temperature quantum field theory, employing  various parameters to represent different physical scenarios. Using these spectral functions, transport coefficients  related to the spin alignment and tensor polarization are obtained,  connecting the spin alignment of the $\phi$ meson to hydrodynamic gradients. Using the standard freeze-out picture within the relativistic hydrodynamic model, we explore how the possible pattern of the spin alignment can be generated using these microscopically calculated spectral functions. We discover that, with certain assumptions, these simple microscopic model-based calculations can produce sizable spin alignments with a sign-flipping behavior in their $p_T$ and centrality dependence, similar to those observed in experiments. 
\end{abstract}
	
\maketitle
	
\section{Introduction}
After the experimental discovery~\cite{STAR:2017ckg} of the predicted global polarization~\cite{Liang:2004ph} of the Lambda hyperon, the spin-related degrees of freedom become a new probe for us to understand the microscopic properties of the hot and dense nuclear matter created in heavy-ion collisions. Although theoretical predictions based on thermal vorticity~\cite{Li:2017slc, Karpenko:2016jyx} successfully describe the experimental observations on the global polarization of the Lambda hyperon, subsequent measurements of the local polarization along the beam direction~\cite{Adam:2019srw} have revealed a sign difference between the experimental results and theoretical predictions based solely on thermal vorticity~\cite{Becattini:2017gcx}. This sign difference between theoretical predictions and  experimental observations is sometimes referred to as  ``spin-sign puzzle"   and has inspired a series of theoretical works aiming to resolve it~\cite{Becattini:2018duy,Liu:2019krs,Li:2020eon,Speranza:2020ilk,Fukushima:2020ucl,Hongo:2021ona,Liu:2020dxg,Liu:2021uhn,Fu:2021pok,Becattini:2021suc,Becattini:2021iol,Wang:2020pej,Ivanov:2020qqe,Lisa:2021zkj,Florkowski:2021wvk,Florkowski:2021xvy,Yi:2021ryh,Lin:2022tma,Kumar:2022ylt,Wu:2022mkr,Fang:2023bbw,Kumar:2023ojl,Hidaka:2023oze,Gao:2023wwo,Wang:2024lis,Sheng:2024pbw,Wang:2024lis,Weickgenannt:2024esg,Fang:2024vds,Dey:2024cwo}. Among all these attempts,  it is worth mentioning the discovery of a new mechanism for generating local polarization—shear-induced polarization~\cite{Liu:2021uhn, Fu:2021pok, Becattini:2021suc,Becattini:2021iol}, which can generate the correct sign in given scenarios, making significant progress in solving the puzzles. 

Besides hyperon polarization, there is another type of polarization—tensor polarization/spin alignment. It has also been predicted to be an observable~\cite{Liang:2004xn} and subsequently confirmed in experiments~\cite{STAR:2008lcm, ALICE:2019aid, STAR:2022fan, ALICE:2022dyy}. Similar to the hyperon polarization case, there are also puzzling phenomena observed for the spin alignment. The values of the spin alignment observed in experiments are found to be orders of magnitude larger than those predicted by previous theories~\cite{Becattini:2007nd, Becattini:2007sr, Xia:2020tyd, Becattini:2022zvf}. Meanwhile, the observed spin alignment exhibits complex $p_T$ and centrality dependence, including sign-flipping behaviors.  There are many attempts aiming to understand this puzzling spin alignment observed in experiments~\cite{Sheng:2019kmk, Muller:2021hpe, Sheng:2022ffb, Sheng:2022wsy,Li:2022vmb,Wagner:2022gza, Wei:2023pdf,Dong:2023cng,Yin:2025gvl}. Among these efforts, the new effects predicted by the  fluctuation-dissipation relation~\cite{Li:2022vmb,Li:2025pef} demonstrate that spin alignment is intrinsically related to the spectral functions of the vector meson in-medium, calling for more detailed studies on this direction.

In heavy-ion collisions, one of the most well-studied in-medium spectral properties is that of the $\rho$ meson. Following the discovery of the excess in dilepton spectra~\cite{CERES:1995vll}, transport approaches~\cite{Li:1995qm,Cassing:1995zv} have attempted to interpret the excess as a manifestation of the dropping mass effects suggested by the sum rule~\cite{Hatsuda:1991ez,Hatsuda:1992bv} and Brown-Rho scaling~\cite{Brown:1991kk}. Around the same time, a competing many-body approach~\cite{Chanfray:1995jgo,Rapp:1997fs} has also been proposed to explain the excess by the broadening and distortions of $\rho$ meson spectral functions caused by many-body interactions. Later, experiments with higher collision energy were conducted at SPS~\cite{NA60:2006ymb}, RHIC~\cite{PHENIX:2009gyd,STAR:2013pwb,PHENIX:2015vek}, and LHC~\cite{ALICE:2018ael} on this topic. To understand these experiments, there are two competing theoretical models based on either the many-body approach~\cite{Rapp:2000pe,vanHees:2006ng,vanHees:2007th,Rapp:2013ema} or the transport model~\cite{Linnyk:2011hz,Linnyk:2011vx,Cassing:2009vt,Bratkovskaya:2011wp}, both of which include the broadening effects of the $\rho$ meson. Besides the $\rho$ meson, the in-medium spectral properties of the heavy quarkonium have been also widely studied. Theoretically, the in-medium spectral functions have been investigated using various ways, including lattice QCD (LQCD)~\cite{Alberico:2006vw,Aarts:2011sm,Ding:2012sp,Aarts:2014cda}, T-matrix approach~\cite{Liu:2016ysz,Liu:2017qah}, and others as reviewed in~\cite{Rothkopf:2019ipj}. Experimentally, the observed  $R_{AA}$ and $v_2$ data of the heavy quarkonium are closely correlated with the dissociation/regeneration rate, binding energies of the quarkonium states in-medium~\cite{Andronic:2024oxz}, which are closely related to the quarkonium spectral properties (widths, mass shifts, etc.). 

For the $\phi$ meson, whose spin alignment has been  measured, there are also studies on its spectral properties. In the early work, the $\phi$ meson spectral function was studied within the thermal field theory for chiral perturbation theory~\cite{Haglin:1994ap,Song:1996gw}.
Subsequently, using the many-body approach, which is mainly developed to study the $\rho$ meson, the $\phi$ meson spectral function is also calculated~\cite{Rapp:2000pe,vanHees:2007th}. Using the forward scattering amplitude, an alternative calculation is carried out~\cite{Vujanovic:2009wr}, including many collisional contributions. There are also proposals that one can obtain information about the spectral properties of the $\phi$ meson from experimental measurements on kaon pairs or dileptons~ \cite{Li:1994cj,Ko:1997kb,Pal:2002aw}, but will be more challenging compared to those of the $\rho$ meson. The lack of the experimental support makes our understanding of the $\phi$ meson spectral properties much less compared to that of $\rho$ meson.  Also, there are much fewer studies on its spectral properties from LQCD than those of quarkonium.

The discovery of the spin alignment's connection to spectral functions of vector mesons provides new opportunities for us to study the in-medium spectral properties of the QCD matter. As demonstrated in our previous work~\cite{Li:2022vmb}, the behavior of the spin alignment is sensitive to the underlying in-medium spectral functions of the $\phi$ meson.  However, a more detailed discussion on how the spectral functions calculated through the microscopic model can affect the spin alignment is still lacking. Especially, there are no microscopic calculations using pure hadronic degrees of freedom for studies on spin alignment of the $\phi$ mesons. 
On the other hand, as proposed in Ref~\cite{Li:2022vmb}, the complex behaviors of the spin alignment observed in experiments are probably related to the fact that in certain experimental conditions, the spin alignment of the $\phi$ meson freezes out mainly in the hadronic phase, while in other conditions, it mainly freezes out in the quark-gluon plasma(QGP) phase. Therefore, a combined study of a theory with purely hadronic degrees of freedom and a theory that can characterize the resonance of the $\phi$ meson in the QGP can help us to further illustrate this picture.

In this work, we will employ the simplest Lagrangian involving the $\phi$ meson from the SU(3) chiral perturbation theory, to study the spin alignment in the hadronic phase. Since the main focus of the work is qualitative features, we will only re-calculate the simplest one-loop diagrams that maintain the Ward-Identity, which has been studied before~\cite{Haglin:1994ap},  with various parameters for our studies on the spin alignments. For the spin alignment in the partonic phase, we will also present one-loop calculations of the quark-meson model shown in our previous work~\cite{Li:2022vmb}.  With these two models, we will discuss how the spin alignments vary in different scenarios and how they are related to the complex behaviors observed in experiments. 

The paper will be organized as follows: in Sec.~\ref{sec_theo}, we will 
discuss the microscopic theory/model based on the hadronic field theory and the quark-meson model, where we will calculate the self-energy diagrams with both frameworks at the one-loop level. Then, in Sec.~\ref{sec_spec}, with the expressions of these self-energy diagrams, we will calculate the corresponding spectral functions and transport coefficients for the spin alignment. In  Sec.~\ref{sec_phenom}, we will discuss the spin alignment based on these calculated microscopic coefficients. In Sec.~\ref{sec_sum}, we will summarize the findings and provide a perspective on the future.
 
\section{Theoretical Formalism}
\label{sec_theo}
\subsection{ Spin alignments and spectral properties}
The tensor polarization and spin alignment can be derived using the linear response theory within the Zubarev response approach~\cite{Li:2025pef}. Some of the content below has also been reported in our previous work~\cite{Li:2022vmb}, and as first discussed in this work, the tensor polarization (for its relation to spin alignment, see Eq.~(\ref{eq_rho-nT})) of vector mesons in a hydrodynamic medium can be expressed as
\begin{align}
	\label{eq_T1full-re}
	\sT^{\mu\nu}
	=&\beta n(\ve_{\vp}) \tilde\Delta^{\langle\mu}_\l\tilde\Delta^{\nu\rangle}_\g \left( \alpha_{{\rm sh}} \xi^{\g\l}  +\a_\text{sp}\xi_p\frac{u^\l u^\g}{-\tilde v^2}
	+\alpha_0\frac{u^\l u^\g}{-\tilde v^2} \right),
\end{align}
where $n(\ve_{\vp})$ is the density distribution of bosons, $\ve_{\vp}=\sqrt{\mathbf{p}^2+m^2}$,  $m$ is the  mass of the vector meson, $\beta = 1/T$ is the inverse temperature,  $u^\mu(x)$ is the  flow velocity, $\xi_{\l\g} \equiv \beta^{-1} \partial_{(\l} (\beta u)_{\g)}$ is the symmetric thermal flow gradient, $\tilde{\Delta}^{\mu\nu}=-g^{\mu\nu}+\tilde{p}^{\mu}\tilde{p}^{\nu}/m^2$ with $\tilde p^\mu = (\ve_{\vp}, \vp)$, $\tilde{\Delta}_{\lambda}^{\langle \mu}\tilde{\Delta}^{\nu \rangle}_{\gamma} \equiv \tilde{\Delta}_{\lambda}^{( \mu}\tilde{\Delta}^{\nu )}_{\gamma}-\tilde{\Delta}^{\mu\nu}\tilde{\Delta}_{\lambda\gamma}/3$, $\tilde{v}^{\mu}=\tilde{\Delta}^{\mu\nu}u_{\nu}$, and $\xi_{p} \equiv (\tilde p^\rho \tilde p^\sigma)\xi_{\rho\sigma}/\ve_{\vp}^2 $. These three coefficients $\alpha_\text{sh}$, $\alpha_\text{sp}$, $\alpha_\text{0}$ are dissipative coefficients closely related to the hydrodynamic medium.
With $a=T,L$ and $d_T=2$ and $d_L=1$, we have
\begin{align}
	\label{eq_spec}
	&A^{\mu\nu}=\frac{1}{\pi} \text{Im} D^{\mu\nu}(\o, \vp)=\sum_{a=L,T}\Delta^{\mu\nu}_{a}A_{a}(\o,\vp),\no\\
	&A_{a}=\frac{1}{\pi}\text{Im }\frac{-1}{p^2-m^2-\Pi_a}, \Pi_a=\frac{-1}{d_a}\D^{\mu\nu}_a\Pi_{\mu\nu}.
\end{align}
The longitudinal and transverse projectors $ \Delta_{T,L} $ are  $ \Delta^{\mu\nu}_{L}=v^\mu v^\nu/(-v^2)$, 
$ \Delta^{\mu\nu}_{T}=\Delta^{\mu\nu}-\Delta^{\mu\nu}_{L}$, where
$v^\mu=\Delta^{\mu\nu}u_\nu$ and $\Delta^{\mu\nu}=-g^{\mu\nu}+p^\m p^\n/p^2$.  The $D^{\m\n}$ is the in-medium propagator. The expressions for the coefficients $\alpha_{\text{sh}}$, $\alpha_{\text{sp}}$, $\alpha_{\text{0}}$ are as follows:
\begin{align}
	\label{eq_ash}
	&\alpha_{\rm sh}= \frac{4\ve_{\vp} \pi}{ \beta n(\ve_{\vp})}\int^\infty_0d\o\frac{\partial n(\o)}{\partial \o}  (\o^2-\ve_{\vp}^2) A^2_{T/L}(\o,\vp),\no\\
	&\a_\text{sp}= \frac{4\ve_{\vp} \pi}{\beta n(\ve_{\vp})}\int^\infty_0d\o\frac{\partial n(\o)}{\partial \o} \ve^2_{\vp} (A_T^2(\o,\vp)-A_L^2(\o,\vp)),\no\\
	&\a_\text{0} = 2\ve_{\vp}\int_0^\infty d\o\, \frac{n(\o)}{n(\ve_{\vp})} \Big{[}(A_L-A_T)-\frac{\D\o^2\tilde{v}^2}{p^2}A_L\Big{]}.
\end{align}
Furthermore, in the quasi-particle limit, the spectral function near the positive frequency pole can be approximately expressed as 
\begin{equation}
	\label{eq_quasiA}
	A_a(\omega,\vp) \approx \frac{1}{2\ve_{\vp}}\frac{1}{\pi} \mathrm{Im} \frac{-1}{\omega - \omega^a_{\vp}+i \Gamma^a_{\vp}/2},
\end{equation}
where $\Gamma_{\vp}^{a} =\text{Im}\Pi_{a}(\o^{a}_{\vp},\vp)/\ve_{\vp} $ is the width, and $\omega^a_{\vp}$ is the in-medium energy/dispersion relation. In the quasi-particle limit, the coefficients $\alpha_{\rm sh}$ and $\alpha_{\rm sp}$ can be expressed as
\begin{align}
	\label{eq_ash2}
	&\alpha_{\rm sh} \approx
	-\frac{2\Delta\ve_{\vp}}{\Gamma_{\vp}}+2\frac{\Delta\ve_{\vp}}{\Gamma_{\vp}}\frac{\Delta\ve_{\vp}}{T}+\frac{\Gamma_{\vp}}{2T} \sim \mathcal O(1),\\
	&\a_\text{sp}
	\approx-\frac{\ve_{\vp}}{\G_{\vp}}\left(\frac{\G^{\D}_{\vp}}{\G_{\vp}}-\frac{\D\ve_{\vp}}{T}\frac{\G^{\D}_{\vp}}{\G_{\vp}}+\frac{\G_{\vp}}{T}\frac{\o^{\D}_{\vp}}{\G_{\vp}}\right).\no
\end{align}
 $\a_0$ does not have a general quasi-particle limit, unlike $\a_{\rm sh}$ and $\a_{\rm sp}$.  However, for some non-analytical self-energy diagrams, it can have an approximate expression $\a_0\approx (\o_{T}-\o_{L})/T$.
In this part, $\G^{\D}_{\vp} \equiv\G^{L}_{\vp}-\G^{T}_{\vp}$ and $\o^{\D}_{\vp} \equiv\o^{L}_{\vp}-\o^{T}_{\vp}$ correspond to the differences in widths and dispersion relations between the longitudinal($L$) and transverse($T$) modes, respectively. 
$\Delta \ve_{\vp}$ and $\G_{\vp}$ are defined as $\Delta \ve_{\vp}=\o_{\vp}^{L/T}-\ve_{\vp}$ and $\G_{\vp}=\G^{L/T}_{\vp}$, where the differences
caused by choosing L/T  are  $\mathcal O(\d_{\text{sp}})$.

\subsection{Spectral properties for the hadronic field theory}
\label{sec_hadronic}
The quantum field of the hadronic degrees of freedom can be mostly described by the structure predicted by the spontaneous breaking of the $SU(3)_{\text{L}}\times SU(3)_{\text{R}}$ chiral symmetry  to $SU(3)_{V}$. Typically, the chiral symmetry in these Lagrangians is realized non-linearly based on the development discussed in\cite{Coleman:1969sm,Callan:1969sn} together with anomalous Wess-Zumino-Witten term\cite{Wess:1971yu}. There are several equivalent ways to generate the effective Lagrangian~\cite{Ecker:1988te,Ecker:1989yg,Meissner:1987ge,Bando:1987br}. For heavy-ion physics, rather than taking the antisymmetric tensor form for the vector boson~\cite{Ecker:1988te,Ecker:1989yg}, we still take the more common vector field form~\cite{Gale:1990pn,Haglin:1994ap,Hohler:2013ena,Hohler:2015iba}. To obtain the minimal Lagrangian for  the study of $\phi$ meson spin alignment in this work, we will follow the Yang-Mills approach in Ref.~\cite{Meissner:1987ge}, which is 
\begin{align}
	\label{eq_L_hadron}
	\mathcal{L}=&|\pd_\mu K|^2+|\pd_\mu K_0|^2+i g_{\phi} \phi^{\mu}(K^*\overset{\leftrightarrow}{\pd}_\mu K+K^*_0\overset{\leftrightarrow}{\pd}_\mu K_0)\no\\
	&+g_{\phi}^2 (|K_0|^2+|K|^{2})\phi^\mu\phi_\mu-\frac{1}{4}\phi_{\mu\nu}\phi^{\mu\nu}+\frac{1}{2}m_\phi ^2\phi^\mu \phi_\mu\no\\
	&-m_K^2 |K|^2-m_{K_0}^2 |K_0|^2,
\end{align}
where the $X\overset{\leftrightarrow}{\pd}_\mu Y=X\pd_\mu Y-(\pd_\mu X) Y$, $g_{\phi}=g/2$, $\phi_{\mu\nu}=\pd_\mu \phi_\nu-\pd_\nu \phi_\mu$, which is the same as those used in Ref.~\cite{Cabrera:2002hc}. Due to the
assumption that the mass of $K_0$ is degenerate with $K$, we
denote both as $m_K$ and use a universal coupling constant $g_\phi$.

In vacuum, the vertices of the above Lagrangian are $ig_\phi (q_\mu+k_\mu)(2\pi)^4\delta^{(4)}(p+q-k),\;2i g_\phi^2g_{\m\n}(2\pi)^4\delta^{(4)}(p+l+q-k)$, for one can use Chapter 61 of the textbook~\cite{Srednicki:2007qs} as a benchmark in the scalar QED field theory. For finite-temperature Feynman rules, we need to make the following modifications in order to obtain the Feynman rules of Matsubara formalism: substitute all real-time energy variables with Matsubara frequencies as $q_0\rightarrow i \o_n$; times all propagators with $i$; divide all vertices by $i$. 

\begin{figure} [!htb]
	\centering
	\includegraphics[width=0.99\columnwidth]{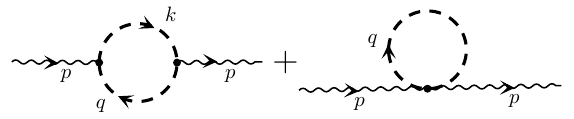}
	\caption{Self-energy diagrams of the hadronic field theory}
	\label{self-energy of hadronic}
\end{figure}

With $q^{\mu}=(i\o_n,\vq)$ and $k=q+p$, the expression of the self-energy diagrams in Fig.~\ref{self-energy of hadronic}  is
\begin{align}
	\Pi^{\mu\nu}(p)=&\frac{2 g_\phi^2}{\b}\sum_{n}\int\frac{d^3 \vq}{(2\pi)^3}\frac{2 q^{\mu}+p^{\mu}}{q^2-m_K^2}\frac{2q^{\nu}+p^{\nu}}{(p+q)^2-m_K^2}\no\\
	&-\frac{4 g_\phi^2}{\b}\sum_{n}\int\frac{d^3 \vq}{(2\pi)^3}\frac{g^{\mu\nu}}{q^2-m_K^2}.
\end{align}
Here considering the contributions of  both $K$ and $K_0$, there exists a whole factor 2 in the above formula.  In order to evaluate the summation, we will transform the summation with a contour integral using the residue theorem. Here we adopt the formula $\beta^{-1}\sum_{n} X(i \o_n)= \pm(2\pi i)^{-1}\oint dz n(z) X(z)$ for a generic function $X$, where the $n(z)=1/(e^{\beta z}\mp1)$
 and the contour is anticlockwise around the imaginary axis. When the poles are $i\omega_n=2i n\pi  /\beta$ , $n(z)=1/(e^{\beta z}-1)$ and when the poles are $i\omega_n=i(2n+1)\pi/\beta$ , $n(z)=1/(e^{\beta z}+1)$. Using the above trick, $\Pi^{\mu\nu}(p)$ will be
\begin{align}
	\label{eq_tofz1}
	&\Pi^{\mu\nu}(p)=2 g_\phi^2  \oint \frac{dz}{2\pi i} n (z)\int\frac{d^3 \vq}{(2\pi)^3} I(z,\vq),\no\\
	&I(z,\vq)=\frac{2 q^{\mu}+p^{\mu}}{q^2-m_K^2}\frac{2q^{\nu}+p^{\nu}}{(p+q)^2-m_K^2}
	-\frac{2g^{\mu\nu}}{q^2-m_K^2}.
\end{align}
Here for the reason that both $K$ and $K_0$ are bosons, $i\omega_n=2in\pi/\beta$ and $n(z)=1/(e^{\beta z}-1)$. And if we choose the contour around the imaginary axis, with a small positive or negative shift, we can explicitly write down the contour integral as
\begin{align}
	\label{eq_vac+me_b}
	\Pi^{\mu\nu}&=2 g_\phi^2\int\frac{d^3 \vq}{(2\pi)^3}\frac{1 }{2\pi i}\big[ \int_{i\infty-\e}^{-i\infty-\e}dz\, n  (z)I(z,\vq)\no\\ 
	&\hspace{1cm}+\int_{-i\infty+\e}^{i \infty+\e}dz\, n  (z)I(z,\vq)\big]\no\\
	&=2 g_\phi^2 \int\frac{d^3 \vq}{(2\pi)^3}\frac{1}{2\pi i} \big[\int_{-i\infty-\e}^{i \infty-\e}dz (1+n  (-z))I(z,\vq)\no\\ 
	&\hspace{1cm}+\int_{-i\infty+\e}^{i \infty+\e}dz\,n (z)I(z,\vq)\big],
\end{align}
where we can split it into two parts:
\begin{align}
&\Pi^{\mu\nu}=\Pi^{\mu\nu}_{\rm vac}+\Pi^{\mu\nu}_{\rm me},\no\\
&\Pi^{\mu\nu}_{\rm me}=-2g^2_{\phi}\sum_{i=L,R}\int \frac{d^3\vq}{(2\pi)^3}\frac{s_i}{2\pi i}\oint_{C_{i}} dz \hspace{0.3em} 
 n(s_{i}z)I(z,\vq),
\end{align}
 where $s_{L/R}=\mp1$. $\Pi^{\mu\nu}_{\rm vac}$ is the one related to the ``1" part in the third line of Eq.~(\ref{eq_vac+me_b}), which will be discussed  later. The $C_{L/R}$ refers to the contour of the left or right half counter-clock circle ( `` $\;\cleftsemicirc$ " or  `` $\;\crightsemicirc$ "). The integrand has the symmetry  under the integral such that $\int d z d^3\vq \;n(z)I (z,\vq)=\int d z d^3\vq \;n(z)I (-z-p_0,-\vq-\vp)$, we can further simplify the  integral for $\Pi^{\mu\nu}_{\rm me}$ as
\begin{align}
	\Pi^{\mu\nu}_{\rm me}
	=&-2 g_\phi^2\int\frac{d^3 \vq}{(2\pi)^3} \frac{1}{2\pi i} \oint_{C_R} dz\,2n  (z)I(z,\vq),
\end{align}
with $\tilde{q}=(\ve_{\vq},\vq)$ where $\ve_{\vq}=\sqrt{\mathbf{q}^2+m_K^2}$. Performing the integral using the residue theorem, we obtain
\begin{align}
	\label{eq_seflme}
 	&\Pi^{\mu\nu}_{\text{me}}(p)=-4 g_\phi^2 \int\frac{d^3 \vq}{(2\pi)^3}\Big\{\frac{1}{4\ve_{\vk}\ve_{\vq}}
 	\Big[n(\ve_{\vq})(2 \tilde{q}^{\mu}+p^{\mu})\no\\
 	&\times(2 \tilde{q}^{\nu}+p^{\nu})\frac{2\ve_{\vk}}{(\tilde{q}+p)^2-m_K^2}+n(\ve_{\vk})(2 \tilde{k}^{\mu}-p^{\mu})(2 \tilde{k}^{\nu}-p^{\nu})\no\\
 	&\times\frac{2\ve_{\vq}}{(\tilde{k}-p)^2-m_K^2}\Big]-2g^{\mu\nu}\frac{n(\ve_{\vq})}{2\ve_{\vq}}\Big\},
 \end{align}
 where $\tilde{k}=(\ve_{\vk},\vk)$ and $\vk=\vp+\vq$.  Performing the change of variables $\vq\rightarrow-(\vq+\vp)$ for the second term in the $[...]$ of Eq.~(\ref{eq_seflme}), then the second term becomes the same as the first term in $[...] $, except that the $\tilde{q}^\m=(\ve_{\vq},\vq)$ in the first term needs to be replaced by  $\tilde{q}^\m=(-\ve_{\vq},\vq)$. Eventually, with $q=(q_0,\vq)$,   Eq.~(\ref{eq_seflme}) can be further simplified as
\begin{align}
	\label{eq_seflme2}
	&\Pi^{\mu\nu}_{\text{me}}(p)=-2 g_\phi^2 \int\frac{d^4 q}{(2\pi)^3}\Big\{2n(\ve_{\vq})\text{sign}(q_0)\rho_K(q_0,\vq)
\no\\
	&\hspace{1.6cm}\times	\Big[\frac{(2 q^{\mu}+p^{\mu})(2 q^{\nu}+p^{\nu})}{(q+p)^2-m_K^2}-g^{\mu\nu}\Big]\Big\},\\
	\label{eq_rhok}
	&\rho_K(q_0,\vq)=\frac{\delta(q_0-\ve_{\vq})-\delta(q_0+\ve_{\vq})}{2\ve_{\vq}},
\end{align}
where sign(x) is the sign function. Note that the angular integral can be calculated analytically, and only a 1D
$q$ integral remains.

If we contract  $\Pi^{\mu\nu}_{\text{me}}(p)$ with $p^\m=(p+q)^\m-q^\m$, we can verify the transversality/Ward-identity of  $\Pi^{\mu\nu}_{\text{me}}(p)$ as follows:
\begin{align}
	\label{eq_ward}
	p_\mu\Pi^{\mu\nu}_{\text{me}}(p)=&-2 g_\phi^2 \int\frac{d^4 q}{(2\pi)^3}\Big\{2n(\ve_{\vq})\text{sign}(q_0)\rho_K(q_0,\vq)
	\no\\
	&\times	[(2 q^{\nu}+p^{\nu})-p^{\nu}]\Big\}=0.
\end{align}
Note that the integral  vanishes due to the fact that the integrand is an odd function with respect to  $q^{\n}$ . In this sense, the transversality is automatically retained at the one-loop level.

For the vacuum term, the numerical result of self-energy diagrams can be evaluated using the dimensional regularization. The expression of the vacuum term is 
\begin{align}
	\Pi^{\mu\nu}_{\text{vac}}(p)
	=&2 g_\phi^2 \int\frac{d^3 \vq}{(2\pi)^3}\frac{1}{2\pi i} \int_{-i \infty}^{i \infty}dz \Big[
	\frac{2 q^{\mu}+p^{\mu}}{q^2-m_K^2}\no\\&\times\frac{2q^{\nu}+p^{\nu}}{(p+q)^2-m_K^2}
	-2g^{\mu\nu}\frac{1}{q^2-m_K^2}\Big].
\end{align}
With $l=q+xp,\,\Delta_K=m_K^2-x(1-x)p^2$, we get
\begin{align}
	\label{eq_vacpi}
	\Pi^{\mu\nu}_{\text{vac}}(p)
	=&2 g_\phi^2\int_{0}^{1}dx \int\frac{d^4 l_E}{(2\pi)^4}\Big[\frac{(1-2x)^2(p^\mu p^\nu-p^2g^{\mu\nu})}{(l_E^2+\Delta_K)^2}\no\\
	&+	\frac{ g^{\mu\nu }(l_E^2+2\Delta_K)}{(l_E^2+\Delta_K)^2}\Big],
\end{align}
where the second term vanishes with the dimensional regularization. The first term can be expressed as $\Pi^{\m\n}_{\text{vac}}=(p^2g^{\mu\nu}-p^{\m} p^{\n})\Pi$, where $\Pi$ is the scalar part of $\Pi^{\m\n}_{\text{vac}}$. After subtracting the infinity, one option of the scalar part of $\Pi^{\m\n}_{\text{vac}}$ is
\begin{align}
	&\Pi=-2\frac{g_\phi^2}{(4\pi)^2}\int_{0}^{1}\mathrm{d}x \hspace{0.3em}(1-2x)^2\ln\bigg(\frac{|m^2_K-(1-x)x m_\phi^2|}{m^2_K-(1-x)x p^2}\bigg).
\end{align}
However, in order to make the pole and residue at the renormalized values, we need to further take account of the finite part of the counterterms $\delta m$ and $\delta_A$.
 The vacuum propagator $D^{\mu\nu
 }(p)$ with counterterms is
 \begin{align}
 	D^{\mu\nu}(p)=\cfrac{g^{\mu\nu}-p^\mu p^\nu/p^2}{p^2(1-\Pi(p^2)-\delta_A)-m_{\phi}^2-\delta_m}\,.
 \end{align}
  With a notice that $\Pi(m^2_\phi)=0$, the renormalization conditions become
 \begin{align}
 	&\text{Re}[m^2_\phi\delta_A+\delta_m]=0,\,\text{Re}[\frac{d [\text{Den}]}{dp^2}|_{m_\phi^2}]=1\no\\
 	&\rightarrow\text{Re}[ \delta_A+m_\phi^2 d\Pi(p)/d p^2|_{p^2=m_\phi^2}]=0.
 \end{align}
Here [Den] represents the denominator of the propagator $D^{\mu\nu}(p)$. Solving the above equations, we get the values of the counterterms as
 \begin{align}
 	\delta_A&=2 \frac{g_\phi^2}{(4\pi)^2}\text{PV}\int_0^{1}dx\frac{(1 - 2 x)^2 (1 - 
 		x) x m_\phi^2}{m_K^2 - (1 - x) x m^2_\phi},\no\\
 	\delta_m&=-m_\phi^2\delta_A ,
 \end{align}
 where PV denotes the principal value integral.
 
 \subsection{Spectral properties in the quark-meson model}
 \label{sec_QMtheo}
To investigate the qualitative features of mesonic degrees of freedom around the phase boundary of the QGP phase and hadronic phase, the quark-meson model is typically employed. In this work, we just take the sector involving the strange quarks and the $\phi$ mesons from the SU(3) quark-meson model~\cite{Zacchi:2015lwa}. For the purpose of this work, we  only calculate the one-loop diagram to demonstrate how spectral properties influence spin alignment qualitatively. The content below has also been reported in our previous work~\cite{Li:2022vmb}.
\begin{figure} [!tb]
	\centering
	\includegraphics[width=0.5\columnwidth]{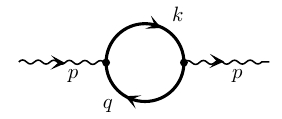}
	\caption{Self-energy diagram of the quark-meson model}
	\label{self-energy of QM}
\end{figure}

 The quark-meson model Lagrangian used in this work can be expressed as
 \begin{align}
 	\mathcal{L}=\,\bar{\psi}(i\slashed{D}-m_s)\psi-\frac{1}{4}\phi^{\mu\nu}\phi_{\mu\nu}+\frac{1}{2}m_{\phi}^2\phi^\mu \phi_\mu,
 \end{align}
 in which $D_{\mu}=\pd_{\mu}-i g_{\phi }\phi_{\mu}$ and $\phi_{\mu\nu} =\partial_\mu \phi_\nu-\partial_\nu \phi_\mu$ with $\phi_{\mu}$ as the $\phi$ meson field. This Lagrangian is the same as QED, except for an extra mass term for the $\phi$ meson. 
 
 Denoting $k=q+p$ and using  Matsubara formalism, the one-loop self-energy diagram can be expressed as
 \begin{align}
 	&\Pi^{\mu\nu}(p)=-\frac{ g_\phi^2}{\b}\sum_{n}\int\frac{d^3 \vq}{(2\pi)^3}\text{Tr}\left\{\gamma^{\mu}\frac{\slashed{q}+m_s}{q^2-m_s^2}\gamma^{\nu}\frac{\slashed{k}+m_s}{k^2-m_s^2}\right\} .
 \end{align}
Similar to those discussed in Sec.\ref{sec_hadronic}, we sum up Matsubara frequencies with the residue theorem and take the trace. Meanwhile, to further simplify the expression,  this technique $(\vq \rightarrow -\vk, \vk\rightarrow-\vq)$ can be used  as discussed in Sec.\ref{sec_hadronic}. After all these procedures, we can also separate the value of the self-energy 
diagram into the vacuum and medium parts. The medium part of the self-energy is
 \begin{align}
 	\Pi^{\mu\nu}_{\rm me}(p)=&-4g_\phi^2\int d q_0 \frac{d^3\vq}{(2\pi)^3}2\,\text{sign}(q_0) n(\ve_{\vq}) \rho_s(q_0,\bm{q})\no\\&\times\frac{q^\mu k^{\nu}+k^\mu q^\nu-g^{\mu\nu}( q\cdot k-m_s^2)}{k^2-m^2_s},
 \end{align}
 with $n(z)=1/(e^{\beta z}+1)$, $\rho_s(q_0,\vq)=(\delta(q_0-\ve_{\vq})-\delta(q_0+\ve_{\vq}))/(2\ve_{\vq})$,  $\ve_{\vq}=\sqrt{m_s^2+\vq^2}$. The $p_0$ should be understood as $p_0\pm i \epsilon$ 
 for analytical continuation. Similar to those in  Sec.\ref{sec_hadronic}, the angular integral can be calculated analytically, leaving only a 1D
 $q$ integral, and  Ward-identity $p_\mu \Pi ^{\mu\nu}_{me}=(k_\mu -q_\mu)\Pi ^{\mu\nu}_{me}=0$ is satisfied.
 
 The vacuum part of the self-energy can be calculated using dimensional regularization. We can get the vacuum part which is $\Pi ^{\mu\nu}_{\rm vac}=(p^2g^{\mu\nu}-p^\mu p^\nu)\Pi$,  where the scalar part $\Pi$ is
 \begin{align}
 	\Pi&=-8\frac{g^2_\phi}{(4\pi)^2}\int_{0}^{1}dx\,x(1-x)\ln\bigg(\frac{|m^2_s-(1-x)x m_\phi^2|}{m^2_s-(1-x)x p^2}\bigg).
 \end{align}
Counterterms $\delta_A$  and $\delta_m$ are tuned to make the vacuum propagator
 \begin{align}
	D^{\mu\nu}(p)=\cfrac{g^{\mu\nu}-p^\mu p^\nu/p^2}{p^2(1-\Pi(p^2)-\delta_A)-m_{\phi}^2-\delta_m}
\end{align}
  have the correct pole and residue, where the expressions of the counterterms are
 \begin{align}
 	\delta_A&=8\frac{g_\phi^2}{(4\pi)^2}\text{PV}\int_0^{1}dx\frac{x(1 -  x) (1 - 
 		x) x m_\phi^2}{m_s^2 - (1 - x) x m^2_\phi},\no\\
 	\delta_m&=-m_\phi^2\delta_A.
 \end{align}
 The PV here denotes the ``Principal Value".

\section{Spectral properties and the transport coefficients}
\label{sec_spec}
Using the results of the self-energy diagrams calculated based on hadronic field theory and the quark-meson model, we will proceed to evaluate these quantities and the corresponding spectral functions numerically; these results will then serve as inputs to obtain the transport coefficients discussed in Ref.~\cite{Li:2022vmb}. We will first calculate those for the hadronic field theory in Sec.~\ref{sec_spec_hadron} and then we will discuss the same physics using the quark-meson model Sec.~\ref{sec_spec_partonic}.
We have benchmarked our numerical results with Ref.~\cite{Haglin:1994ap} for the hadronic field theory and Ref.~\cite{Dong:2023cng} for the quark-meson model. Using the same setups and parameters as in these references, our results are consistent with those in the literature. Also, it should be noted that a similar study for the $\rho$ meson using hadronic field theory was released shortly before this work~\cite{Yin:2025gvl}.

\subsection{Hadronic field theory}
\label{sec_spec_hadron}
\begin{figure} [!htb]
	\centering
	\includegraphics[width=0.99\columnwidth]{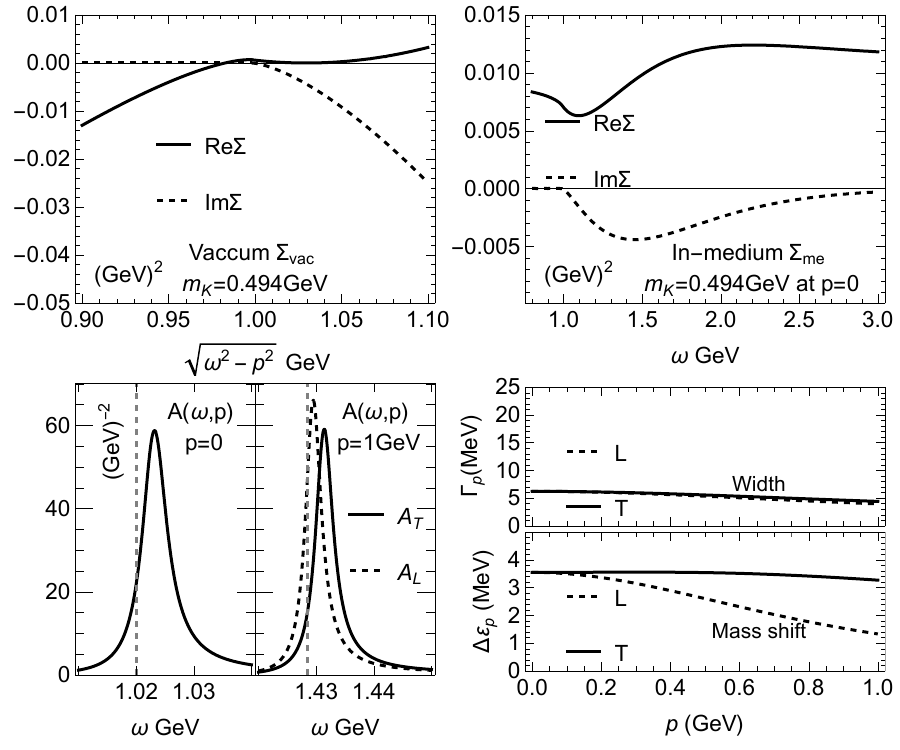}
	\vspace{-0.7cm}
	\caption{Spectral properties with $m_K=0.494$~GeV: first,  vacuum self-energy; second, in-medium self-energy with $p=0$; third, transverse ($T$) and longitudinal ($L$) parts of the spectral function at $p=\{0,1\}~$GeV; fourth, the weighted widths and mass shifts for ``$T$" and ``$L$" modes.   
	}
	\label{fig_hfspec1}
\end{figure}

\begin{figure} [!htb]
	\centering
	\includegraphics[width=0.99\columnwidth]{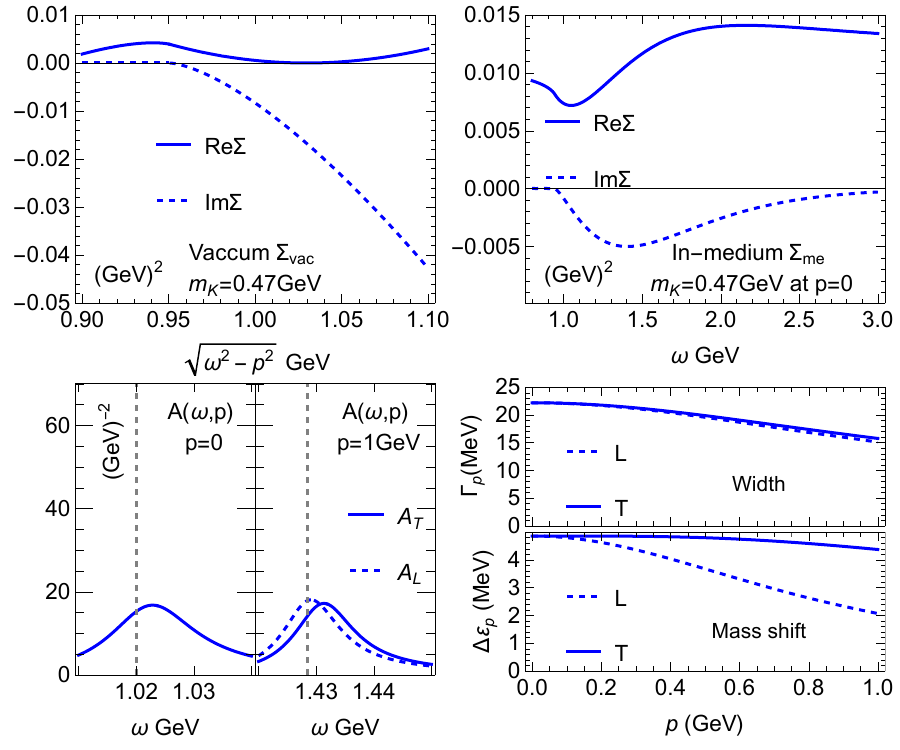}
	\vspace{-0.7cm}
	\caption{Spectral properties with $m_K=0.47$~GeV: first,  vacuum self-energy; second, in-medium self-energy with $p=0$; third, transverse ($T$) and longitudinal ($L$) parts of the spectral function at $p=\{0,1\}~$GeV; fourth, the weighted widths and mass shifts for ``$T$" and ``$L$" modes.   
	}
	\label{fig_hfspec2}
\end{figure}

With the self-energy diagrams of the $\phi$ meson calculated in Sec.~\ref{sec_hadronic}, we plot them in various ways as shown in Fig.~\ref{fig_hfspec1} and Fig.~\ref{fig_hfspec2}. The coupling constant is chosen as $g_\phi=4.55$~\cite{Haglin:1994ap}, which matches the $\phi\rightarrow K\bar{K}$ branching ratio of the total $\phi$. However, in Ref.~\cite{Haglin:1994ap}, it does not consider the interaction term $\phi\rightarrow K_0\bar{K_0}$ predicted by the chiral Lagrangian, as shown in Eq.~(\ref{eq_L_hadron}), but it is discussed in Ref.~\cite{Cabrera:2002hc}. Also, as reported in Ref.~\cite{Haglin:1994ap}, the contribution of the Wess-Zumino-Witten term is small and we neglect it at the moment. 
\begin{figure} [!htb]
	\centering
	\includegraphics[width=0.99\columnwidth]{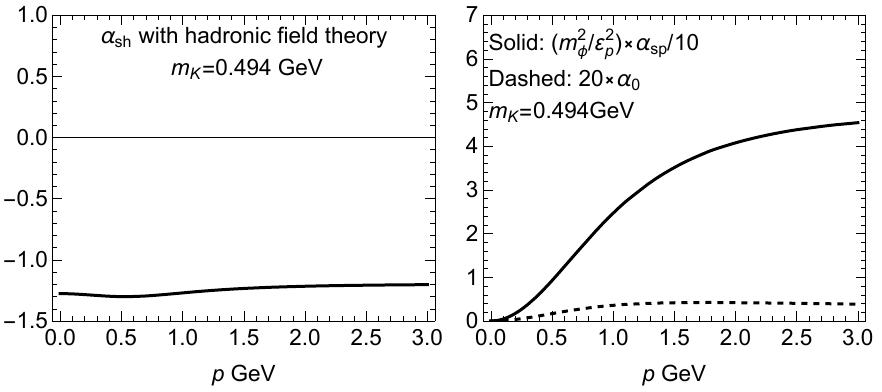}
	\vspace{-0.7cm}
	\caption{The transport coefficients for the tensor polarization: $\a_{\rm sh}$(left); $(m^2_{\phi}/\ve^2_{\vp}) \a_{\rm sp}/10$ and $20\a_0$ (right).   
	}
	\label{fig_hfco1}
\end{figure}
In the medium, the mass and width of the kaon will be modified, with an energy shift (at rest) of the order of 10 MeV~\cite{Faessler:2002qb}. However, naively including the width will mess up the Ward-identity in Eq.~(\ref{eq_ward}). For this reason, we only include this in-medium mass effect of the kaons. To illustrate the effect of this mass shift, we take $m_K=\{0.494, 0.47\}$~GeV in  our calculations. 

For spectral functions deviating from the pure Lorentzian form, there are various definitions for mass shifts and widths. To best suit the approximation in Eq.~(\ref{eq_ash2}), we will adopt the following definitions for the mass shift and width used in the plots:
\begin{align}
	&\D \ve_{\vp}=\frac{\int d\o (\o-\ve_{\vp})A_{a}^2(\o,\vp)}{\int{d\o A^2_{a}(\o,\vp)}},\no\\
	&\G_a=2\sqrt{\frac{\int d\o (\o-\ve_{\vp}-\D \ve_{\vp})^2A^2_{a}(\o,\vp)}{\int{d\o A^2_{a}(\o,\vp)}}}.
\end{align}
The motivation to use the above definitions originates from the form of Eq.~(\ref{eq_ash}).

In Fig.~\ref{fig_hfspec1} and Fig.~\ref{fig_hfco1}, we demonstrate the spectral properties of the $\phi$ meson and the corresponding transport coefficients for spin alignment (see Ref.~\cite{Li:2022vmb} for its definition), with the vacuum  kaon mass that $m_K=0.494$~GeV. For the reason that $\phi$ meson
mass $1.02$~GeV is so close to two times of the kaon mass ($2\times0.494$~GeV), the width is only around 4~MeV, which is just around its vacuum width. On the other hand, we notice that the medium generates a mass shift of a few MeV. At the same time,  the split of the in-medium mass between the longitudinal and transverse mass is significant.

If we take the formalism and the approximation in Ref.~\cite{Li:2022vmb}, the transport coefficients $\a_{\rm sh}$, $\a_{\rm sp}$ and $\a_{ 0}$ in Eq.~(\ref{eq_ash}) are plotted in Fig.~\ref{fig_hfco1}. The general trend of the coefficients can be understood using Eq.~(\ref{eq_ash2}). Since the width is much smaller than the temperature, the term $\G_{\vp}/T$ is negligible and the dominant contribution for $\a_{\rm sh}$  is  $-2\D\ve_{\vp}/\G_{\vp}$, which is negative since the energy shift is positive. Also, due to the large factor $\ve_{\vp}/\G_{\vp}$, $\a_{\rm sp}$ is very large and will contribute significantly to spin alignment. However, we should note that the width of the spectral function is around $~5$ MeV, corresponding to the relaxation time  around 40~fm. The duration of hadronic phase might be too short to allow for the spin alignment to approach equilibrium with these hadronic field Lagrangians.

\begin{figure} [!htb]
	\centering
	\includegraphics[width=0.99\columnwidth]{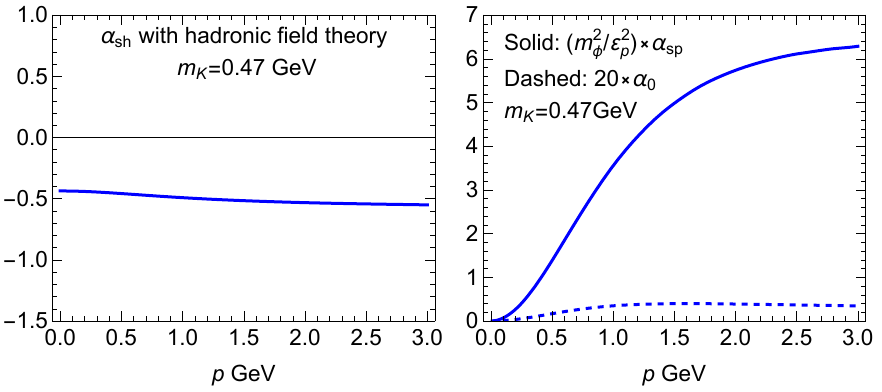}
	\vspace{-0.7cm}
	\caption{The transport coefficients for the tensor polarization: $\a_{\rm sh}$(left); $ (m^2_{\phi}/\ve^2_{\vp}) \a_{\rm sp}$ and $20\a_0$ (right).   
	}
	\label{fig_hfco2}
\end{figure}

However, as mentioned before, the mass of kaons in the medium could be smaller than its vacuum value. If we take $m_K=0.47 $~GeV, the spectral and transport properties of the $\phi$ meson are shown in Fig.~\ref{fig_hfspec2} and Fig.~\ref{fig_hfco2}.

As shown, due to the decrease in mass, the mass of the $\phi$ meson is significantly larger than the two kaon thresholds, allowing a  larger phase space for the interactions and generating a much larger width that is around 20~MeV. This corresponds to a relaxation time around 10 fm, which is of the same order of magnitude as the lifetime of the hadronic phase in central heavy-ion collisions. However, since the mass shift is positive and this width is  small compared to the temperature, $\a_{\rm sh}$ is still negative. $\a_{\rm sp}$ becomes much smaller compared to those in Fig.~\ref{fig_hfco1}, since  $\ve_{\vp}/\G_{\vp}$ gets smaller due to the larger width.



\subsection{Quark-meson model}
 \label{sec_spec_partonic}
\begin{figure} [!htb]
 	\centering
 	\includegraphics[width=0.99\columnwidth]{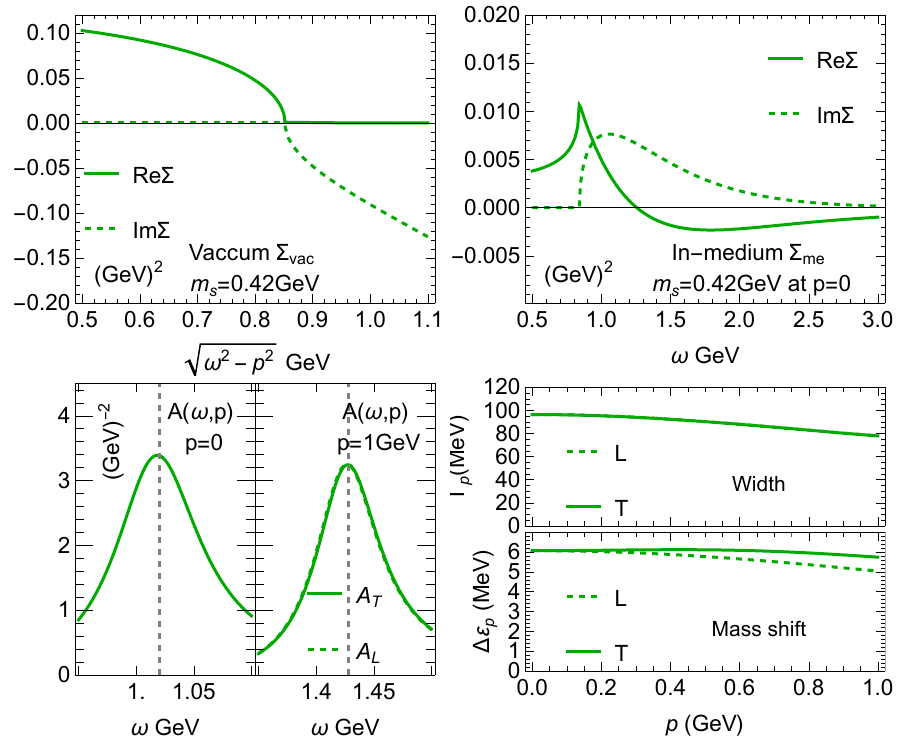}
 	\vspace{-0.7cm}
 	\caption{Spectral properties with $m_s=0.42$~GeV: first,  vacuum self-energy; second, in-medium self-energy with $p=0$; third, transverse ($T$) and longitudinal ($L$) parts of the spectral function at $p=\{0,1\}~$GeV; fourth, the weighted widths and mass shifts for ``$T$" and ``$L$" modes.   
 	}
 	\label{fig_qmspec1}
 \end{figure}

\begin{figure} [!htb]
 	\centering
 	\includegraphics[width=0.99\columnwidth]{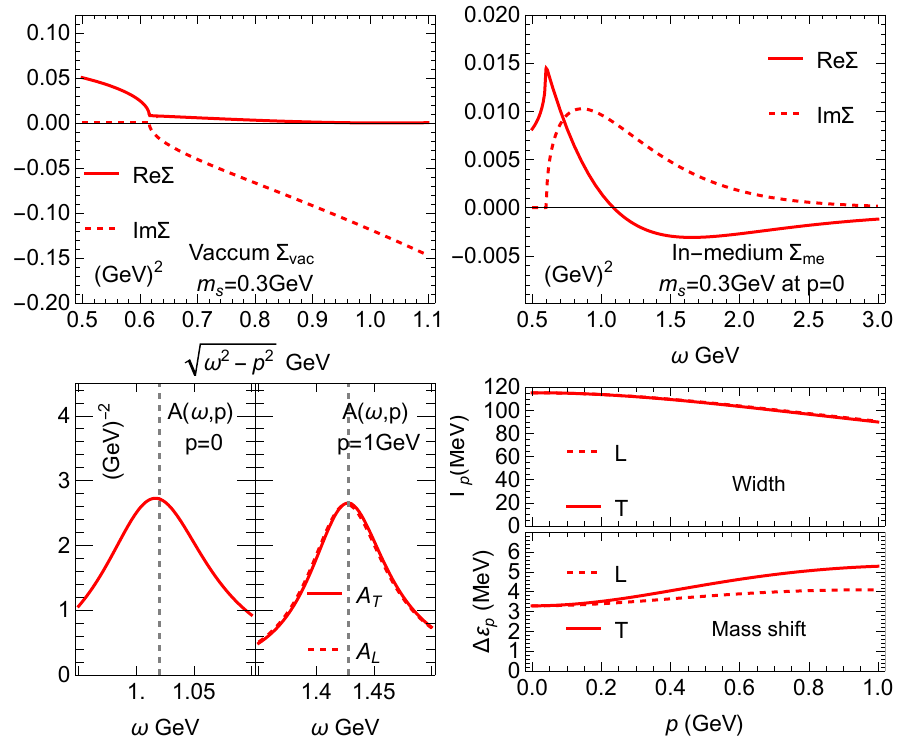}
 	\vspace{-0.7cm}
 	\caption{Spectral properties with $m_s=0.3$~GeV: first,  vacuum self-energy; second, in-medium self-energy with $p=0$; third, transverse ($T$) and longitudinal ($L$) parts of the spectral function at $p=\{0,1\}~$GeV; fourth, the weighted widths and mass shifts for ``$T$" and ``$L$" modes.   
 	}
 	\label{fig_qmspec2}
 \end{figure}

Using the value of the self-energy diagram of the $\phi$ meson from Sec.~\ref{sec_QMtheo}, we plot the results in various ways, as shown in Fig.~\ref{fig_qmspec1} and Fig.~\ref{fig_qmspec2}. Unlike the hadronic field theory, there is no ``vacuum reference"  of  the coupling constant for quark-meson model. In the literature, $g$ is a parameter, which is tuned for various phenomenological purposes~\cite{Zacchi:2015lwa,Dong:2023cng}, with a value of order $1$ being generally reasonable. For our phenomenological purpose, we select $g_{\phi}=2.2$, which allows us to better demonstrate the qualitative features of the theory. For the mass of the  strange quark in the medium,  there is no conclusive answer for what should be used. Here we select two representative masses namely $m_s=0.3$~GeV and $m_s=0.42$~GeV to illustrate the qualitative features of the theory, which is larger than the bare mass of the strange quark and smaller than the quasi-particle mass used in other approaches~\cite{Liu:2017qah}.
 \begin{figure} [!htb]
 	\centering
 	\includegraphics[width=0.99\columnwidth]{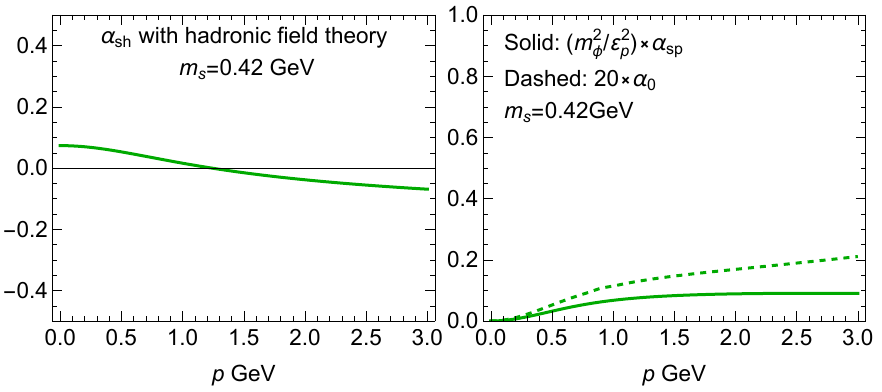}
 	\vspace{-0.7cm}
 	\caption{The transport coefficients for the tensor polarization: $\a_{\rm sh}$(left); $(m_{\phi}^2/\ve_{\vp}^2)\a_{\rm sp} $ and $20\a_0$ (right).   
 	}
 	\label{fig_qmco1}
 \end{figure}
 
 With these setups, we first calculate the spectral properties of the $\phi$ meson using the $m_s$=0.42~GeV. As shown in Fig.~\ref{fig_qmspec1}, the width is around 100~MeV, which is much larger compared to that calculated using the hadronic field theory in Sec.~\ref{sec_spec_hadron}. This occurs because the hadronic field theory is coupled with the derivative which significantly suppresses the width. This resulting large width causes the third term in Eq.~(\ref{eq_ash2}) to dominate, leading to positive values of $\alpha_{\rm sh}$ at low momentum. However, at higher momentum, both the width and energy shift will decrease. This will make the third term suppress much faster than the first term, since the effects of decrease $\D\ve_{\vp}$ and $\G_{\vp}$ will cancel each other in the first term in  Eq.~(\ref{eq_ash2}), leading to a negative $\a_{\rm sh}$ at higher momentum.  As the width and the ratio $\G_{\vp}/\D \ve_{\vp}$ increase, the enhancement factor $\a_{\rm sp}$ becomes much smaller compared to those in Sec.~\ref{sec_spec_hadron}.

 \begin{figure} [!htb]
 	\centering
 	\includegraphics[width=0.99\columnwidth]{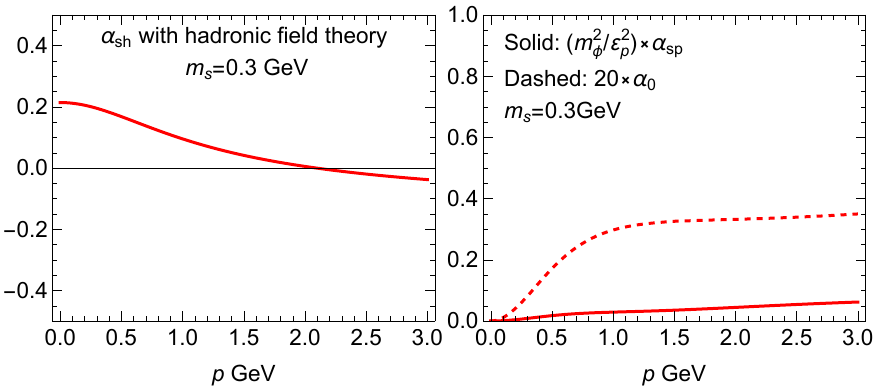}
 	\vspace{-0.7cm}
 	\caption{The transport coefficients for the tensor polarization: $\a_{\rm sh}$(left); $(m_{\phi}^2/\ve_{\vp}^2)\a_{\rm sp}$ and $20\a_0$ (right).   
 	}
 	\label{fig_qmco2}
 \end{figure}
\begin{figure*} [!htb]
	\centering
	\includegraphics[width=1.99\columnwidth]{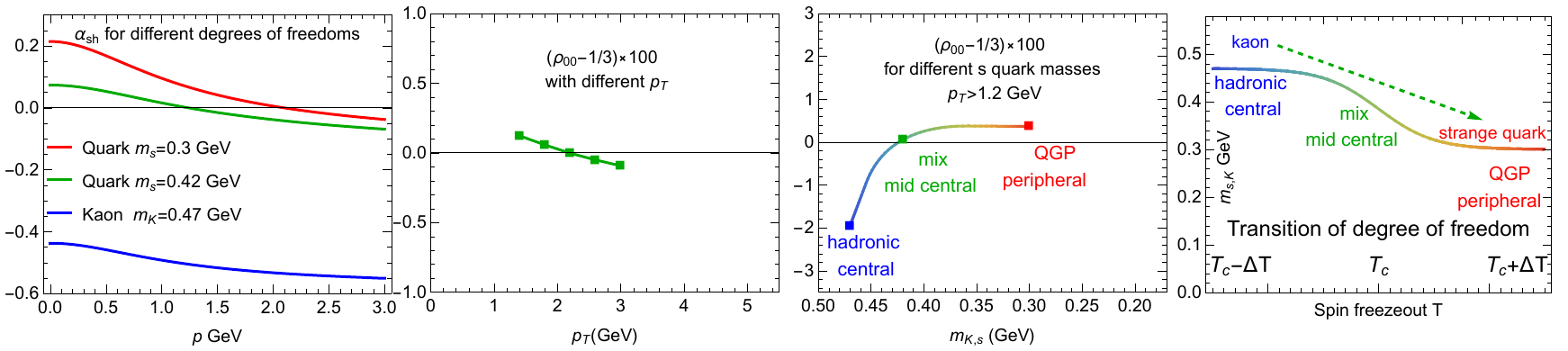}
	\vspace{-0.1cm}
	\caption{First panel: $\a_{\rm sh}$ coefficients for three scenarios; second panel: the $p_T$ dependence for the spin alignment with quark meson model $m_s=0.42$~GeV; third panel: using the mass dependence to mimic the ``centrality" of the spin alignment; fourth panel:illustration of the mapping between mass dependence, freeze-out temperature, and the ``centrality".
	}
	\label{fig_align}
\end{figure*}
When we choose a smaller mass  $m_s=0.3$~GeV for the strange quark, the enlarged phase space can continue to increase the width as shown in Fig.~\ref{fig_qmspec2}. Meanwhile, the energy shifts become smaller. These two effects make $\a_{\rm sh}$ increase positively compared to those in Fig.~\ref{fig_qmco1}. The momentum dependence of the coefficients and the magnitude of  $\a_{\rm sp}$ still follow the same physics as described in Fig.~\ref{fig_qmco1}. As shown, although the width is relatively large, $\a_{\rm sh}$ is not so large, due to cancellation effects of the positive energy shifts from the ``radiative" interactions.  If we have included the ``collisional" effects in the QGP phase, we can create a ``negative" mass shift (see Ref~\cite{Liu:2017qah}), which makes $\a_{\rm sh}$ larger. A generic case study was conducted in our previous work~\cite{Li:2022vmb}.

\section{Phenomenological Analyses for spin alignment}
\label{sec_phenom}
For the phenomenological analyses, except for the coefficients, we will use the same setups as those in our previous short paper~\cite{Li:2022vmb} and we will quote the essential physical concepts discussed in that work in order to make the current work more self-contained~\cite{Li:2022vmb}.

Following Refs~\cite{Becattini:2019ntv,Fu:2021pok,Becattini:2021iol}, we will accept the commonly used freeze-out assumption in this work. Meanwhile, we also adopt the physical picture that $\phi$ mesons carrying the spin alignment are formed in the late stage of the QGP, as discussed in several works~\cite{Shuryak:2004tx,Liu:2017qah}. However, the freezeout of the spin alignment can occur at different phases (QGP, mixed, or hadronic phases) for different evolution paths of the fireball, depending on the conditions of the collisions, such as beam energy, centrality, etc., which might be related to the complex behaviors of the spin alignment as functions of these conditions observed in the experiments.



The standard Cooper-Fry-like formula is employed to characterize the freeze-out of spin alignment. In this formalism, the $\delta \rho_{00}$ and tensor polarization $\sT^{\mu\nu}$ are related to the formula as
\begin{align}
	\label{eq_rho-nT}
	\delta \rho_{00}(\hat{n}_{\text{pr}},\vp)=\frac{\int d\Sigma^{\l}p_{\l}\, \sT^{\mu\nu}(x,\vp)\hn_{\mu}(\vp)\hn_{\nu}(\vp)}{d\Sigma^{\l}p_{\l} 3 n(\ve_u)}\,.
\end{align}
The $\Sigma^\lambda$ is the hyper-surface under the freeze-out condition, and $\hat n^\mu$ is an abbreviation of $\epsilon^\mu_{s=0}$, which is associated with the 3-dimensional unit vector $\hat n_{\text{pr}}$ by a Lorentz boost ($\hat{n}^{\mu}=[\Lambda(\mathbf p).\hat{n}_{\text{pr}}]^{\mu}$), where $\hat{n}_{\text{pr}}$ is the polarization vector $\epsilon^\mu_{s=0,\text{pr}}$ in the rest frame of the particle. In this work, we only consider the case that  $\hat{n}_{\text{pr}}$ is in out-plane ($\hat y$) direction. We employed the hydrodynamic profile based on CLVisc~\cite{Pang:2016igs}, which is used in our previous work~\cite{Fu:2021pok} .  The freeze-out temperature of the profile is 157~MeV.  Meanwhile, the profile uses the AMPT initial condition~\cite{Fu:2021pok} for Au-Au collisions at mid-centrality with the impact parameter being around 9~fm.

 In this paper, the spin alignments of the $\phi$ meson have been evaluated with the coefficients $\a_{\rm sh}$, $\a_{\rm sp}$, $\a_{\rm 0}$ that are calculated with either the hadronic field theory or the quark-meson model. We only choose the $m_K=0.47$~GeV case for the hadronic field theory, since the width of the other case is too small for the $\phi$ meson to have enough time to become equilibrated. 

The results are shown in Fig.~\ref{fig_align}, in which we first organize all three coefficients. The $\a_{\rm sh}$ coefficient lies in the first panel. The second panel illustrates the $p_T$ dependence of spin alignment in the case for the quark-meson model with $m_s=0.42$~GeV, and all contributions of $\a_{\rm sh}$, $\a_{\rm sp}$, $\a_{\rm 0}$ considered. Nevertheless, the dominate effects come from $\a_{\rm sh}$ and the $p_T$ dependence of spin alignment in this case and they can be understood through the green curve, which represents the property of the $\a_{\rm sh}$ coefficients in the first panel.

The centrality dependence is illustrated in the third panel in Fig.~\ref{fig_align}, for which we need the mapping shown in the fourth panel: ``$m_K=0.47$ GeV $\leftrightarrow 0\textendash20\%$'', ``$m_s=0.42$ GeV $\leftrightarrow 20\textendash40\%$'', and ``$m_s=0.3$ GeV $\leftrightarrow 40\textendash60\%$''. The reason for this mapping has already been discussed in Ref~\cite{Li:2022vmb}, and here we will summarize the main idea of it. First, we should notice that the kinetic freezeout temperature of the hadrons decreases under the conditions of more central collisions and higher beam energies~\cite{STAR:2017sal} and the central or higher beam energy collisions have a longer lifetime of the hadronic phase~\cite{Wu:2020zbx}, allowing the hadrons' spectra to re-equilibrate in the hadronic phase with lower temperatures. For the freezeout of spin alignment, the physics should be similar. With an equilibrated spin alignment generated in the QGP phase, if the hadronic phase is short, which is the case for peripheral collisions, then there is not enough time for it to re-equilibrate in the hadronic phase, and it will retain its spin alignment from the QGP phase. On the other hand, with a long hadronic phase in central collisions, it can re-equilibrate in the hadronic phase. Therefore, for different centralities, the effective degrees of freedom and interactions at the freezeout might change from a quark-level Lagrangian to a pure hadronic Lagrangian, in which the s-quark in the loop becomes the kaons. 

As explained in Ref~\cite{Li:2022vmb}, the hydrodynamic profiles for all centralities and for all freezeout temperatures within a certain reasonable range will have a positive spin alignment if the coefficient is positive.
Therefore, the main physics behind the sign-flipping originates from the sign-flipping of the coefficients depending on whether the spin freezes out in the hadronic phase or the QGP phase (or a different freezeout temperature). To focus on the main physics, we use a single hydrodynamic profile for all centralities in this qualitative study. More comprehensive and realistic studies will be carried out later.
Also, for those concerned about whether or not a larger positive $a_{\rm sh}$ is possible and the beam energy dependence of the spin alignment, more discussions can be found in our previous work~\cite{Li:2022vmb}, and more comprehensive studies are in our plans.
\section{Summary and perspective}
\label{sec_sum}
In this work, we employ the thermal field theory to calculate spectral functions of the $\phi$ meson in either a hadronic Lagrangian based on chiral perturbation theory or a quark-meson model Lagrangian. 
With various parameters, we obtain several sets of spectral functions and their corresponding transport coefficients for the tensor polarization and the spin alignment. These results are then employed in the phenomenological study of spin alignment using relativistic hydrodynamic profiles.  

We demonstrate that different underlying interactions (hadronic and partonic) can generate distinct spectral functions and coefficients. More specifically, the one-loop results of the hadronic field theory based on chiral perturbation theory have a tendency to generate smaller widths,  relatively larger energy shifts (considering the small width) and larger energy splitting, resulting in a negative $\a_{\rm sh}$ coefficient and finally a negative spin alignment. 
In contrast, the quark-meson model tends to generate larger widths, relatively small energy shifts, and smaller energy splittings, resulting in positive $\a_{\rm sh}$ and positive spin alignments. Meanwhile, the competition between the effects of the negative energy-shift-over-width term and those of the positive width-over-temperature term leads to the sign-flipping behaviors in the $p_T$-dependent spin alignment.
Assuming the mapping between the centrality and freeze-out with different masses and degrees of freedom, we can generate a sign-flipping ``centrality" dependence of the spin alignment.

However, the current discussion of the spectral properties is just for illustrative purposes. We just use the simplest theoretical setups to demonstrate that different underlying interactions can generate rich physics about spectral functions that might lead to rich behaviors of the spin alignment observed in experiments. 
In the future, we hope for a more conclusive answer to the spin alignment problem. We should make more realistic calculations for the $\phi$ meson spectral functions, such as including various collisional interactions in the hadronic medium, which has already been studied in many contexts~\cite{Vujanovic:2009wr,vanHees:2007th} . Also, we might employ some non-perturbative calculations~\cite{Liu:2017qah,Fu:2019hdw} for the spectral properties.

\acknowledgments
\textbf{Acknowledgments}
This research is supported by NSFC No. 12205090 and Fundamental Research Funds for the Central Universities.
\bibliography{ref}

\begin{thebibliography}{109}%
\makeatletter
\providecommand \@ifxundefined [1]{%
 \@ifx{#1\undefined}
}%
\providecommand \@ifnum [1]{%
 \ifnum #1\expandafter \@firstoftwo
 \else \expandafter \@secondoftwo
 \fi
}%
\providecommand \@ifx [1]{%
 \ifx #1\expandafter \@firstoftwo
 \else \expandafter \@secondoftwo
 \fi
}%
\providecommand \natexlab [1]{#1}%
\providecommand \enquote  [1]{``#1''}%
\providecommand \bibnamefont  [1]{#1}%
\providecommand \bibfnamefont [1]{#1}%
\providecommand \citenamefont [1]{#1}%
\providecommand \href@noop [0]{\@secondoftwo}%
\providecommand \href [0]{\begingroup \@sanitize@url \@href}%
\providecommand \@href[1]{\@@startlink{#1}\@@href}%
\providecommand \@@href[1]{\endgroup#1\@@endlink}%
\providecommand \@sanitize@url [0]{\catcode `\\12\catcode `\$12\catcode
  `\&12\catcode `\#12\catcode `\^12\catcode `\_12\catcode `\%12\relax}%
\providecommand \@@startlink[1]{}%
\providecommand \@@endlink[0]{}%
\providecommand \url  [0]{\begingroup\@sanitize@url \@url }%
\providecommand \@url [1]{\endgroup\@href {#1}{\urlprefix }}%
\providecommand \urlprefix  [0]{URL }%
\providecommand \Eprint [0]{\href }%
\providecommand \doibase [0]{http://dx.doi.org/}%
\providecommand \selectlanguage [0]{\@gobble}%
\providecommand \bibinfo  [0]{\@secondoftwo}%
\providecommand \bibfield  [0]{\@secondoftwo}%
\providecommand \translation [1]{[#1]}%
\providecommand \BibitemOpen [0]{}%
\providecommand \bibitemStop [0]{}%
\providecommand \bibitemNoStop [0]{.\EOS\space}%
\providecommand \EOS [0]{\spacefactor3000\relax}%
\providecommand \BibitemShut  [1]{\csname bibitem#1\endcsname}%
\let\auto@bib@innerbib\@empty
\bibitem [{\citenamefont {Adamczyk}\ \emph
  {et~al.}(2017{\natexlab{a}})\citenamefont {Adamczyk} \emph
  {et~al.}}]{STAR:2017ckg}%
  \BibitemOpen
  \bibfield  {author} {\bibinfo {author} {\bibfnamefont {L.}~\bibnamefont
  {Adamczyk}} \emph {et~al.} (\bibinfo {collaboration} {STAR}),\ }\href
  {\doibase 10.1038/nature23004} {\bibfield  {journal} {\bibinfo  {journal}
  {Nature}\ }\textbf {\bibinfo {volume} {548}},\ \bibinfo {pages} {62}
  (\bibinfo {year} {2017}{\natexlab{a}})}\BibitemShut {NoStop}%
\bibitem [{\citenamefont {Liang}\ and\ \citenamefont
  {Wang}(2005{\natexlab{a}})}]{Liang:2004ph}%
  \BibitemOpen
  \bibfield  {author} {\bibinfo {author} {\bibfnamefont {Z.-T.}\ \bibnamefont
  {Liang}}\ and\ \bibinfo {author} {\bibfnamefont {X.-N.}\ \bibnamefont
  {Wang}},\ }\href {\doibase 10.1103/PhysRevLett.94.102301,
  10.1103/PhysRevLett.96.039901} {\bibfield  {journal} {\bibinfo  {journal}
  {Phys. Rev. Lett.}\ }\textbf {\bibinfo {volume} {94}},\ \bibinfo {pages}
  {102301} (\bibinfo {year} {2005}{\natexlab{a}})},\ \bibinfo {note} {[Erratum:
  Phys. Rev. Lett.96,039901(2006)]}\BibitemShut {NoStop}%
\bibitem [{\citenamefont {Li}\ \emph {et~al.}(2017)\citenamefont {Li},
  \citenamefont {Pang}, \citenamefont {Wang},\ and\ \citenamefont
  {Xia}}]{Li:2017slc}%
  \BibitemOpen
  \bibfield  {author} {\bibinfo {author} {\bibfnamefont {H.}~\bibnamefont
  {Li}}, \bibinfo {author} {\bibfnamefont {L.-G.}\ \bibnamefont {Pang}},
  \bibinfo {author} {\bibfnamefont {Q.}~\bibnamefont {Wang}}, \ and\ \bibinfo
  {author} {\bibfnamefont {X.-L.}\ \bibnamefont {Xia}},\ }\href {\doibase
  10.1103/PhysRevC.96.054908} {\bibfield  {journal} {\bibinfo  {journal} {Phys.
  Rev. C}\ }\textbf {\bibinfo {volume} {96}},\ \bibinfo {pages} {054908}
  (\bibinfo {year} {2017})},\ \Eprint {http://arxiv.org/abs/1704.01507}
  {arXiv:1704.01507 [nucl-th]} \BibitemShut {NoStop}%
\bibitem [{\citenamefont {Karpenko}\ and\ \citenamefont
  {Becattini}(2017)}]{Karpenko:2016jyx}%
  \BibitemOpen
  \bibfield  {author} {\bibinfo {author} {\bibfnamefont {I.}~\bibnamefont
  {Karpenko}}\ and\ \bibinfo {author} {\bibfnamefont {F.}~\bibnamefont
  {Becattini}},\ }\href {\doibase 10.1140/epjc/s10052-017-4765-1} {\bibfield
  {journal} {\bibinfo  {journal} {Eur. Phys. J. C}\ }\textbf {\bibinfo {volume}
  {77}},\ \bibinfo {pages} {213} (\bibinfo {year} {2017})},\ \Eprint
  {http://arxiv.org/abs/1610.04717} {arXiv:1610.04717 [nucl-th]} \BibitemShut
  {NoStop}%
\bibitem [{\citenamefont {Adam}\ \emph {et~al.}(2019)\citenamefont {Adam} \emph
  {et~al.}}]{Adam:2019srw}%
  \BibitemOpen
  \bibfield  {author} {\bibinfo {author} {\bibfnamefont {J.}~\bibnamefont
  {Adam}} \emph {et~al.} (\bibinfo {collaboration} {STAR}),\ }\href {\doibase
  10.1103/PhysRevLett.123.132301} {\bibfield  {journal} {\bibinfo  {journal}
  {Phys. Rev. Lett.}\ }\textbf {\bibinfo {volume} {123}},\ \bibinfo {pages}
  {132301} (\bibinfo {year} {2019})},\ \Eprint
  {http://arxiv.org/abs/1905.11917} {arXiv:1905.11917 [nucl-ex]} \BibitemShut
  {NoStop}%
\bibitem [{\citenamefont {Becattini}\ and\ \citenamefont
  {Karpenko}(2018)}]{Becattini:2017gcx}%
  \BibitemOpen
  \bibfield  {author} {\bibinfo {author} {\bibfnamefont {F.}~\bibnamefont
  {Becattini}}\ and\ \bibinfo {author} {\bibfnamefont {I.}~\bibnamefont
  {Karpenko}},\ }\href {\doibase 10.1103/PhysRevLett.120.012302} {\bibfield
  {journal} {\bibinfo  {journal} {Phys. Rev. Lett.}\ }\textbf {\bibinfo
  {volume} {120}},\ \bibinfo {pages} {012302} (\bibinfo {year}
  {2018})}\BibitemShut {NoStop}%
\bibitem [{\citenamefont {Becattini}\ \emph
  {et~al.}(2019{\natexlab{a}})\citenamefont {Becattini}, \citenamefont
  {Florkowski},\ and\ \citenamefont {Speranza}}]{Becattini:2018duy}%
  \BibitemOpen
  \bibfield  {author} {\bibinfo {author} {\bibfnamefont {F.}~\bibnamefont
  {Becattini}}, \bibinfo {author} {\bibfnamefont {W.}~\bibnamefont
  {Florkowski}}, \ and\ \bibinfo {author} {\bibfnamefont {E.}~\bibnamefont
  {Speranza}},\ }\href {\doibase 10.1016/j.physletb.2018.12.016} {\bibfield
  {journal} {\bibinfo  {journal} {Phys. Lett. B}\ }\textbf {\bibinfo {volume}
  {789}},\ \bibinfo {pages} {419} (\bibinfo {year} {2019}{\natexlab{a}})},\
  \Eprint {http://arxiv.org/abs/1807.10994} {arXiv:1807.10994 [hep-th]}
  \BibitemShut {NoStop}%
\bibitem [{\citenamefont {Liu}\ \emph {et~al.}(2020)\citenamefont {Liu},
  \citenamefont {Sun},\ and\ \citenamefont {Ko}}]{Liu:2019krs}%
  \BibitemOpen
  \bibfield  {author} {\bibinfo {author} {\bibfnamefont {S.~Y.~F.}\
  \bibnamefont {Liu}}, \bibinfo {author} {\bibfnamefont {Y.}~\bibnamefont
  {Sun}}, \ and\ \bibinfo {author} {\bibfnamefont {C.~M.}\ \bibnamefont {Ko}},\
  }\href {\doibase 10.1103/PhysRevLett.125.062301} {\bibfield  {journal}
  {\bibinfo  {journal} {Phys. Rev. Lett.}\ }\textbf {\bibinfo {volume} {125}},\
  \bibinfo {pages} {062301} (\bibinfo {year} {2020})},\ \Eprint
  {http://arxiv.org/abs/1910.06774} {arXiv:1910.06774 [nucl-th]} \BibitemShut
  {NoStop}%
\bibitem [{\citenamefont {Li}\ \emph {et~al.}(2021)\citenamefont {Li},
  \citenamefont {Stephanov},\ and\ \citenamefont {Yee}}]{Li:2020eon}%
  \BibitemOpen
  \bibfield  {author} {\bibinfo {author} {\bibfnamefont {S.}~\bibnamefont
  {Li}}, \bibinfo {author} {\bibfnamefont {M.~A.}\ \bibnamefont {Stephanov}}, \
  and\ \bibinfo {author} {\bibfnamefont {H.-U.}\ \bibnamefont {Yee}},\ }\href
  {\doibase 10.1103/PhysRevLett.127.082302} {\bibfield  {journal} {\bibinfo
  {journal} {Phys. Rev. Lett.}\ }\textbf {\bibinfo {volume} {127}},\ \bibinfo
  {pages} {082302} (\bibinfo {year} {2021})},\ \Eprint
  {http://arxiv.org/abs/2011.12318} {arXiv:2011.12318 [hep-th]} \BibitemShut
  {NoStop}%
\bibitem [{\citenamefont {Speranza}\ and\ \citenamefont
  {Weickgenannt}(2021)}]{Speranza:2020ilk}%
  \BibitemOpen
  \bibfield  {author} {\bibinfo {author} {\bibfnamefont {E.}~\bibnamefont
  {Speranza}}\ and\ \bibinfo {author} {\bibfnamefont {N.}~\bibnamefont
  {Weickgenannt}},\ }\href {\doibase 10.1140/epja/s10050-021-00455-2}
  {\bibfield  {journal} {\bibinfo  {journal} {Eur. Phys. J. A}\ }\textbf
  {\bibinfo {volume} {57}},\ \bibinfo {pages} {155} (\bibinfo {year} {2021})},\
  \Eprint {http://arxiv.org/abs/2007.00138} {arXiv:2007.00138 [nucl-th]}
  \BibitemShut {NoStop}%
\bibitem [{\citenamefont {Fukushima}\ and\ \citenamefont
  {Pu}(2021)}]{Fukushima:2020ucl}%
  \BibitemOpen
  \bibfield  {author} {\bibinfo {author} {\bibfnamefont {K.}~\bibnamefont
  {Fukushima}}\ and\ \bibinfo {author} {\bibfnamefont {S.}~\bibnamefont {Pu}},\
  }\href {\doibase 10.1016/j.physletb.2021.136346} {\bibfield  {journal}
  {\bibinfo  {journal} {Phys. Lett. B}\ }\textbf {\bibinfo {volume} {817}},\
  \bibinfo {pages} {136346} (\bibinfo {year} {2021})},\ \Eprint
  {http://arxiv.org/abs/2010.01608} {arXiv:2010.01608 [hep-th]} \BibitemShut
  {NoStop}%
\bibitem [{\citenamefont {Hongo}\ \emph {et~al.}(2021)\citenamefont {Hongo},
  \citenamefont {Huang}, \citenamefont {Kaminski}, \citenamefont {Stephanov},\
  and\ \citenamefont {Yee}}]{Hongo:2021ona}%
  \BibitemOpen
  \bibfield  {author} {\bibinfo {author} {\bibfnamefont {M.}~\bibnamefont
  {Hongo}}, \bibinfo {author} {\bibfnamefont {X.-G.}\ \bibnamefont {Huang}},
  \bibinfo {author} {\bibfnamefont {M.}~\bibnamefont {Kaminski}}, \bibinfo
  {author} {\bibfnamefont {M.}~\bibnamefont {Stephanov}}, \ and\ \bibinfo
  {author} {\bibfnamefont {H.-U.}\ \bibnamefont {Yee}},\ }\href {\doibase
  10.1007/JHEP11(2021)150} {\bibfield  {journal} {\bibinfo  {journal} {JHEP}\
  }\textbf {\bibinfo {volume} {11}},\ \bibinfo {pages} {150} (\bibinfo {year}
  {2021})},\ \Eprint {http://arxiv.org/abs/2107.14231} {arXiv:2107.14231
  [hep-th]} \BibitemShut {NoStop}%
\bibitem [{\citenamefont {Liu}\ and\ \citenamefont
  {Yin}(2021{\natexlab{a}})}]{Liu:2020dxg}%
  \BibitemOpen
  \bibfield  {author} {\bibinfo {author} {\bibfnamefont {S.~Y.~F.}\
  \bibnamefont {Liu}}\ and\ \bibinfo {author} {\bibfnamefont {Y.}~\bibnamefont
  {Yin}},\ }\href {\doibase 10.1103/PhysRevD.104.054043} {\bibfield  {journal}
  {\bibinfo  {journal} {Phys. Rev. D}\ }\textbf {\bibinfo {volume} {104}},\
  \bibinfo {pages} {054043} (\bibinfo {year} {2021}{\natexlab{a}})},\ \Eprint
  {http://arxiv.org/abs/2006.12421} {arXiv:2006.12421 [nucl-th]} \BibitemShut
  {NoStop}%
\bibitem [{\citenamefont {Liu}\ and\ \citenamefont
  {Yin}(2021{\natexlab{b}})}]{Liu:2021uhn}%
  \BibitemOpen
  \bibfield  {author} {\bibinfo {author} {\bibfnamefont {S.~Y.~F.}\
  \bibnamefont {Liu}}\ and\ \bibinfo {author} {\bibfnamefont {Y.}~\bibnamefont
  {Yin}},\ }\href {\doibase 10.1007/JHEP07(2021)188} {\bibfield  {journal}
  {\bibinfo  {journal} {JHEP}\ }\textbf {\bibinfo {volume} {07}},\ \bibinfo
  {pages} {188} (\bibinfo {year} {2021}{\natexlab{b}})},\ \Eprint
  {http://arxiv.org/abs/2103.09200} {arXiv:2103.09200 [hep-ph]} \BibitemShut
  {NoStop}%
\bibitem [{\citenamefont {Fu}\ \emph {et~al.}(2021)\citenamefont {Fu},
  \citenamefont {Liu}, \citenamefont {Pang}, \citenamefont {Song},\ and\
  \citenamefont {Yin}}]{Fu:2021pok}%
  \BibitemOpen
  \bibfield  {author} {\bibinfo {author} {\bibfnamefont {B.}~\bibnamefont
  {Fu}}, \bibinfo {author} {\bibfnamefont {S.~Y.~F.}\ \bibnamefont {Liu}},
  \bibinfo {author} {\bibfnamefont {L.}~\bibnamefont {Pang}}, \bibinfo {author}
  {\bibfnamefont {H.}~\bibnamefont {Song}}, \ and\ \bibinfo {author}
  {\bibfnamefont {Y.}~\bibnamefont {Yin}},\ }\href {\doibase
  10.1103/PhysRevLett.127.142301} {\bibfield  {journal} {\bibinfo  {journal}
  {Phys. Rev. Lett.}\ }\textbf {\bibinfo {volume} {127}},\ \bibinfo {pages}
  {142301} (\bibinfo {year} {2021})},\ \Eprint
  {http://arxiv.org/abs/2103.10403} {arXiv:2103.10403 [hep-ph]} \BibitemShut
  {NoStop}%
\bibitem [{\citenamefont {Becattini}\ \emph
  {et~al.}(2021{\natexlab{a}})\citenamefont {Becattini}, \citenamefont
  {Buzzegoli},\ and\ \citenamefont {Palermo}}]{Becattini:2021suc}%
  \BibitemOpen
  \bibfield  {author} {\bibinfo {author} {\bibfnamefont {F.}~\bibnamefont
  {Becattini}}, \bibinfo {author} {\bibfnamefont {M.}~\bibnamefont
  {Buzzegoli}}, \ and\ \bibinfo {author} {\bibfnamefont {A.}~\bibnamefont
  {Palermo}},\ }\href {\doibase 10.1016/j.physletb.2021.136519} {\bibfield
  {journal} {\bibinfo  {journal} {Phys. Lett. B}\ }\textbf {\bibinfo {volume}
  {820}},\ \bibinfo {pages} {136519} (\bibinfo {year} {2021}{\natexlab{a}})},\
  \Eprint {http://arxiv.org/abs/2103.10917} {arXiv:2103.10917 [nucl-th]}
  \BibitemShut {NoStop}%
\bibitem [{\citenamefont {Becattini}\ \emph
  {et~al.}(2021{\natexlab{b}})\citenamefont {Becattini}, \citenamefont
  {Buzzegoli}, \citenamefont {Inghirami}, \citenamefont {Karpenko},\ and\
  \citenamefont {Palermo}}]{Becattini:2021iol}%
  \BibitemOpen
  \bibfield  {author} {\bibinfo {author} {\bibfnamefont {F.}~\bibnamefont
  {Becattini}}, \bibinfo {author} {\bibfnamefont {M.}~\bibnamefont
  {Buzzegoli}}, \bibinfo {author} {\bibfnamefont {G.}~\bibnamefont
  {Inghirami}}, \bibinfo {author} {\bibfnamefont {I.}~\bibnamefont {Karpenko}},
  \ and\ \bibinfo {author} {\bibfnamefont {A.}~\bibnamefont {Palermo}},\ }\href
  {\doibase 10.1103/PhysRevLett.127.272302} {\bibfield  {journal} {\bibinfo
  {journal} {Phys. Rev. Lett.}\ }\textbf {\bibinfo {volume} {127}},\ \bibinfo
  {pages} {272302} (\bibinfo {year} {2021}{\natexlab{b}})},\ \Eprint
  {http://arxiv.org/abs/2103.14621} {arXiv:2103.14621 [nucl-th]} \BibitemShut
  {NoStop}%
\bibitem [{\citenamefont {Wang}\ \emph {et~al.}(2021)\citenamefont {Wang},
  \citenamefont {Guo},\ and\ \citenamefont {Zhuang}}]{Wang:2020pej}%
  \BibitemOpen
  \bibfield  {author} {\bibinfo {author} {\bibfnamefont {Z.}~\bibnamefont
  {Wang}}, \bibinfo {author} {\bibfnamefont {X.}~\bibnamefont {Guo}}, \ and\
  \bibinfo {author} {\bibfnamefont {P.}~\bibnamefont {Zhuang}},\ }\href
  {\doibase 10.1140/epjc/s10052-021-09586-8} {\bibfield  {journal} {\bibinfo
  {journal} {Eur. Phys. J. C}\ }\textbf {\bibinfo {volume} {81}},\ \bibinfo
  {pages} {799} (\bibinfo {year} {2021})},\ \Eprint
  {http://arxiv.org/abs/2009.10930} {arXiv:2009.10930 [hep-th]} \BibitemShut
  {NoStop}%
\bibitem [{\citenamefont {Ivanov}(2020)}]{Ivanov:2020qqe}%
  \BibitemOpen
  \bibfield  {author} {\bibinfo {author} {\bibfnamefont {Y.~B.}\ \bibnamefont
  {Ivanov}},\ }\href {\doibase 10.1103/PhysRevC.102.044904} {\bibfield
  {journal} {\bibinfo  {journal} {Phys. Rev. C}\ }\textbf {\bibinfo {volume}
  {102}},\ \bibinfo {pages} {044904} (\bibinfo {year} {2020})},\ \Eprint
  {http://arxiv.org/abs/2006.14328} {arXiv:2006.14328 [nucl-th]} \BibitemShut
  {NoStop}%
\bibitem [{\citenamefont {Lisa}\ \emph {et~al.}(2021)\citenamefont {Lisa},
  \citenamefont {Barbon}, \citenamefont {Chinellato}, \citenamefont {Serenone},
  \citenamefont {Shen}, \citenamefont {Takahashi},\ and\ \citenamefont
  {Torrieri}}]{Lisa:2021zkj}%
  \BibitemOpen
  \bibfield  {author} {\bibinfo {author} {\bibfnamefont {M.~A.}\ \bibnamefont
  {Lisa}}, \bibinfo {author} {\bibfnamefont {J.~a. G.~P.}\ \bibnamefont
  {Barbon}}, \bibinfo {author} {\bibfnamefont {D.~D.}\ \bibnamefont
  {Chinellato}}, \bibinfo {author} {\bibfnamefont {W.~M.}\ \bibnamefont
  {Serenone}}, \bibinfo {author} {\bibfnamefont {C.}~\bibnamefont {Shen}},
  \bibinfo {author} {\bibfnamefont {J.}~\bibnamefont {Takahashi}}, \ and\
  \bibinfo {author} {\bibfnamefont {G.}~\bibnamefont {Torrieri}},\ }\href
  {\doibase 10.1103/PhysRevC.104.L011901} {\bibfield  {journal} {\bibinfo
  {journal} {Phys. Rev. C}\ }\textbf {\bibinfo {volume} {104}},\ \bibinfo
  {pages} {011901} (\bibinfo {year} {2021})},\ \Eprint
  {http://arxiv.org/abs/2101.10872} {arXiv:2101.10872 [hep-ph]} \BibitemShut
  {NoStop}%
\bibitem [{\citenamefont {Florkowski}\ \emph
  {et~al.}(2022{\natexlab{a}})\citenamefont {Florkowski}, \citenamefont
  {Ryblewski}, \citenamefont {Singh},\ and\ \citenamefont
  {Sophys}}]{Florkowski:2021wvk}%
  \BibitemOpen
  \bibfield  {author} {\bibinfo {author} {\bibfnamefont {W.}~\bibnamefont
  {Florkowski}}, \bibinfo {author} {\bibfnamefont {R.}~\bibnamefont
  {Ryblewski}}, \bibinfo {author} {\bibfnamefont {R.}~\bibnamefont {Singh}}, \
  and\ \bibinfo {author} {\bibfnamefont {G.}~\bibnamefont {Sophys}},\ }\href
  {\doibase 10.1103/PhysRevD.105.054007} {\bibfield  {journal} {\bibinfo
  {journal} {Phys. Rev. D}\ }\textbf {\bibinfo {volume} {105}},\ \bibinfo
  {pages} {054007} (\bibinfo {year} {2022}{\natexlab{a}})},\ \Eprint
  {http://arxiv.org/abs/2112.01856} {arXiv:2112.01856 [hep-ph]} \BibitemShut
  {NoStop}%
\bibitem [{\citenamefont {Florkowski}\ \emph
  {et~al.}(2022{\natexlab{b}})\citenamefont {Florkowski}, \citenamefont
  {Kumar}, \citenamefont {Mazeliauskas},\ and\ \citenamefont
  {Ryblewski}}]{Florkowski:2021xvy}%
  \BibitemOpen
  \bibfield  {author} {\bibinfo {author} {\bibfnamefont {W.}~\bibnamefont
  {Florkowski}}, \bibinfo {author} {\bibfnamefont {A.}~\bibnamefont {Kumar}},
  \bibinfo {author} {\bibfnamefont {A.}~\bibnamefont {Mazeliauskas}}, \ and\
  \bibinfo {author} {\bibfnamefont {R.}~\bibnamefont {Ryblewski}},\ }\href
  {\doibase 10.1103/PhysRevC.105.064901} {\bibfield  {journal} {\bibinfo
  {journal} {Phys. Rev. C}\ }\textbf {\bibinfo {volume} {105}},\ \bibinfo
  {pages} {064901} (\bibinfo {year} {2022}{\natexlab{b}})},\ \Eprint
  {http://arxiv.org/abs/2112.02799} {arXiv:2112.02799 [hep-ph]} \BibitemShut
  {NoStop}%
\bibitem [{\citenamefont {Yi}\ \emph {et~al.}(2021)\citenamefont {Yi},
  \citenamefont {Pu},\ and\ \citenamefont {Yang}}]{Yi:2021ryh}%
  \BibitemOpen
  \bibfield  {author} {\bibinfo {author} {\bibfnamefont {C.}~\bibnamefont
  {Yi}}, \bibinfo {author} {\bibfnamefont {S.}~\bibnamefont {Pu}}, \ and\
  \bibinfo {author} {\bibfnamefont {D.-L.}\ \bibnamefont {Yang}},\ }\href
  {\doibase 10.1103/PhysRevC.104.064901} {\bibfield  {journal} {\bibinfo
  {journal} {Phys. Rev. C}\ }\textbf {\bibinfo {volume} {104}},\ \bibinfo
  {pages} {064901} (\bibinfo {year} {2021})},\ \Eprint
  {http://arxiv.org/abs/2106.00238} {arXiv:2106.00238 [hep-ph]} \BibitemShut
  {NoStop}%
\bibitem [{\citenamefont {Lin}\ and\ \citenamefont {Wang}(2022)}]{Lin:2022tma}%
  \BibitemOpen
  \bibfield  {author} {\bibinfo {author} {\bibfnamefont {S.}~\bibnamefont
  {Lin}}\ and\ \bibinfo {author} {\bibfnamefont {Z.}~\bibnamefont {Wang}},\
  }\href {\doibase 10.1007/JHEP12(2022)030} {\bibfield  {journal} {\bibinfo
  {journal} {JHEP}\ }\textbf {\bibinfo {volume} {12}},\ \bibinfo {pages} {030}
  (\bibinfo {year} {2022})},\ \Eprint {http://arxiv.org/abs/2206.12573}
  {arXiv:2206.12573 [hep-ph]} \BibitemShut {NoStop}%
\bibitem [{\citenamefont {Kumar}\ \emph {et~al.}(2023)\citenamefont {Kumar},
  \citenamefont {M\"uller},\ and\ \citenamefont {Yang}}]{Kumar:2022ylt}%
  \BibitemOpen
  \bibfield  {author} {\bibinfo {author} {\bibfnamefont {A.}~\bibnamefont
  {Kumar}}, \bibinfo {author} {\bibfnamefont {B.}~\bibnamefont {M\"uller}}, \
  and\ \bibinfo {author} {\bibfnamefont {D.-L.}\ \bibnamefont {Yang}},\ }\href
  {\doibase 10.1103/PhysRevD.107.076025} {\bibfield  {journal} {\bibinfo
  {journal} {Phys. Rev. D}\ }\textbf {\bibinfo {volume} {107}},\ \bibinfo
  {pages} {076025} (\bibinfo {year} {2023})},\ \Eprint
  {http://arxiv.org/abs/2212.13354} {arXiv:2212.13354 [nucl-th]} \BibitemShut
  {NoStop}%
\bibitem [{\citenamefont {Wu}\ \emph {et~al.}(2022)\citenamefont {Wu},
  \citenamefont {Yi}, \citenamefont {Qin},\ and\ \citenamefont
  {Pu}}]{Wu:2022mkr}%
  \BibitemOpen
  \bibfield  {author} {\bibinfo {author} {\bibfnamefont {X.-Y.}\ \bibnamefont
  {Wu}}, \bibinfo {author} {\bibfnamefont {C.}~\bibnamefont {Yi}}, \bibinfo
  {author} {\bibfnamefont {G.-Y.}\ \bibnamefont {Qin}}, \ and\ \bibinfo
  {author} {\bibfnamefont {S.}~\bibnamefont {Pu}},\ }\href {\doibase
  10.1103/PhysRevC.105.064909} {\bibfield  {journal} {\bibinfo  {journal}
  {Phys. Rev. C}\ }\textbf {\bibinfo {volume} {105}},\ \bibinfo {pages}
  {064909} (\bibinfo {year} {2022})},\ \Eprint
  {http://arxiv.org/abs/2204.02218} {arXiv:2204.02218 [hep-ph]} \BibitemShut
  {NoStop}%
\bibitem [{\citenamefont {Fang}\ \emph {et~al.}(2024)\citenamefont {Fang},
  \citenamefont {Pu},\ and\ \citenamefont {Yang}}]{Fang:2023bbw}%
  \BibitemOpen
  \bibfield  {author} {\bibinfo {author} {\bibfnamefont {S.}~\bibnamefont
  {Fang}}, \bibinfo {author} {\bibfnamefont {S.}~\bibnamefont {Pu}}, \ and\
  \bibinfo {author} {\bibfnamefont {D.-L.}\ \bibnamefont {Yang}},\ }\href
  {\doibase 10.1103/PhysRevD.109.034034} {\bibfield  {journal} {\bibinfo
  {journal} {Phys. Rev. D}\ }\textbf {\bibinfo {volume} {109}},\ \bibinfo
  {pages} {034034} (\bibinfo {year} {2024})},\ \Eprint
  {http://arxiv.org/abs/2311.15197} {arXiv:2311.15197 [hep-ph]} \BibitemShut
  {NoStop}%
\bibitem [{\citenamefont {Kumar}\ \emph {et~al.}(2024)\citenamefont {Kumar},
  \citenamefont {Yang},\ and\ \citenamefont {Gubler}}]{Kumar:2023ojl}%
  \BibitemOpen
  \bibfield  {author} {\bibinfo {author} {\bibfnamefont {A.}~\bibnamefont
  {Kumar}}, \bibinfo {author} {\bibfnamefont {D.-L.}\ \bibnamefont {Yang}}, \
  and\ \bibinfo {author} {\bibfnamefont {P.}~\bibnamefont {Gubler}},\ }\href
  {\doibase 10.1103/PhysRevD.109.054038} {\bibfield  {journal} {\bibinfo
  {journal} {Phys. Rev. D}\ }\textbf {\bibinfo {volume} {109}},\ \bibinfo
  {pages} {054038} (\bibinfo {year} {2024})},\ \Eprint
  {http://arxiv.org/abs/2312.16900} {arXiv:2312.16900 [nucl-th]} \BibitemShut
  {NoStop}%
\bibitem [{\citenamefont {Hidaka}\ \emph {et~al.}(2024)\citenamefont {Hidaka},
  \citenamefont {Hongo}, \citenamefont {Stephanov},\ and\ \citenamefont
  {Yee}}]{Hidaka:2023oze}%
  \BibitemOpen
  \bibfield  {author} {\bibinfo {author} {\bibfnamefont {Y.}~\bibnamefont
  {Hidaka}}, \bibinfo {author} {\bibfnamefont {M.}~\bibnamefont {Hongo}},
  \bibinfo {author} {\bibfnamefont {M.~A.}\ \bibnamefont {Stephanov}}, \ and\
  \bibinfo {author} {\bibfnamefont {H.-U.}\ \bibnamefont {Yee}},\ }\href
  {\doibase 10.1103/PhysRevC.109.054909} {\bibfield  {journal} {\bibinfo
  {journal} {Phys. Rev. C}\ }\textbf {\bibinfo {volume} {109}},\ \bibinfo
  {pages} {054909} (\bibinfo {year} {2024})},\ \Eprint
  {http://arxiv.org/abs/2312.08266} {arXiv:2312.08266 [hep-ph]} \BibitemShut
  {NoStop}%
\bibitem [{\citenamefont {Gao}\ and\ \citenamefont {Yang}(2024)}]{Gao:2023wwo}%
  \BibitemOpen
  \bibfield  {author} {\bibinfo {author} {\bibfnamefont {J.-H.}\ \bibnamefont
  {Gao}}\ and\ \bibinfo {author} {\bibfnamefont {S.-Z.}\ \bibnamefont {Yang}},\
  }\href {\doibase 10.1088/1674-1137/ad2363} {\bibfield  {journal} {\bibinfo
  {journal} {Chin. Phys. C}\ }\textbf {\bibinfo {volume} {48}},\ \bibinfo
  {pages} {053114} (\bibinfo {year} {2024})},\ \Eprint
  {http://arxiv.org/abs/2308.16616} {arXiv:2308.16616 [hep-ph]} \BibitemShut
  {NoStop}%
\bibitem [{\citenamefont {Wang}\ and\ \citenamefont
  {Lin}(2024)}]{Wang:2024lis}%
  \BibitemOpen
  \bibfield  {author} {\bibinfo {author} {\bibfnamefont {Z.}~\bibnamefont
  {Wang}}\ and\ \bibinfo {author} {\bibfnamefont {S.}~\bibnamefont {Lin}},\
  }\href@noop {} {\  (\bibinfo {year} {2024})},\ \Eprint
  {http://arxiv.org/abs/2411.19550} {arXiv:2411.19550 [hep-ph]} \BibitemShut
  {NoStop}%
\bibitem [{\citenamefont {Sheng}\ \emph {et~al.}(2024)\citenamefont {Sheng},
  \citenamefont {Becattini}, \citenamefont {Huang},\ and\ \citenamefont
  {Zhang}}]{Sheng:2024pbw}%
  \BibitemOpen
  \bibfield  {author} {\bibinfo {author} {\bibfnamefont {X.-L.}\ \bibnamefont
  {Sheng}}, \bibinfo {author} {\bibfnamefont {F.}~\bibnamefont {Becattini}},
  \bibinfo {author} {\bibfnamefont {X.-G.}\ \bibnamefont {Huang}}, \ and\
  \bibinfo {author} {\bibfnamefont {Z.-H.}\ \bibnamefont {Zhang}},\ }\href
  {\doibase 10.1103/PhysRevC.110.064908} {\bibfield  {journal} {\bibinfo
  {journal} {Phys. Rev. C}\ }\textbf {\bibinfo {volume} {110}},\ \bibinfo
  {pages} {064908} (\bibinfo {year} {2024})},\ \Eprint
  {http://arxiv.org/abs/2407.12130} {arXiv:2407.12130 [hep-th]} \BibitemShut
  {NoStop}%
\bibitem [{\citenamefont {Weickgenannt}\ and\ \citenamefont
  {Blaizot}(2024)}]{Weickgenannt:2024esg}%
  \BibitemOpen
  \bibfield  {author} {\bibinfo {author} {\bibfnamefont {N.}~\bibnamefont
  {Weickgenannt}}\ and\ \bibinfo {author} {\bibfnamefont {J.-P.}\ \bibnamefont
  {Blaizot}},\ }\href@noop {} {\  (\bibinfo {year} {2024})},\ \Eprint
  {http://arxiv.org/abs/2412.05733} {arXiv:2412.05733 [nucl-th]} \BibitemShut
  {NoStop}%
\bibitem [{\citenamefont {Fang}\ and\ \citenamefont {Pu}(2024)}]{Fang:2024vds}%
  \BibitemOpen
  \bibfield  {author} {\bibinfo {author} {\bibfnamefont {S.}~\bibnamefont
  {Fang}}\ and\ \bibinfo {author} {\bibfnamefont {S.}~\bibnamefont {Pu}},\
  }\href@noop {} {\  (\bibinfo {year} {2024})},\ \Eprint
  {http://arxiv.org/abs/2408.09877} {arXiv:2408.09877 [hep-ph]} \BibitemShut
  {NoStop}%
\bibitem [{\citenamefont {Dey}\ and\ \citenamefont {Das}(2024)}]{Dey:2024cwo}%
  \BibitemOpen
  \bibfield  {author} {\bibinfo {author} {\bibfnamefont {S.}~\bibnamefont
  {Dey}}\ and\ \bibinfo {author} {\bibfnamefont {A.}~\bibnamefont {Das}},\
  }\href@noop {} {\  (\bibinfo {year} {2024})},\ \Eprint
  {http://arxiv.org/abs/2410.04141} {arXiv:2410.04141 [nucl-th]} \BibitemShut
  {NoStop}%
\bibitem [{\citenamefont {Liang}\ and\ \citenamefont
  {Wang}(2005{\natexlab{b}})}]{Liang:2004xn}%
  \BibitemOpen
  \bibfield  {author} {\bibinfo {author} {\bibfnamefont {Z.-T.}\ \bibnamefont
  {Liang}}\ and\ \bibinfo {author} {\bibfnamefont {X.-N.}\ \bibnamefont
  {Wang}},\ }\href {\doibase 10.1016/j.physletb.2005.09.060} {\bibfield
  {journal} {\bibinfo  {journal} {Phys. Lett. B}\ }\textbf {\bibinfo {volume}
  {629}},\ \bibinfo {pages} {20} (\bibinfo {year} {2005}{\natexlab{b}})},\
  \Eprint {http://arxiv.org/abs/nucl-th/0411101} {arXiv:nucl-th/0411101}
  \BibitemShut {NoStop}%
\bibitem [{\citenamefont {Abelev}\ \emph {et~al.}(2008)\citenamefont {Abelev}
  \emph {et~al.}}]{STAR:2008lcm}%
  \BibitemOpen
  \bibfield  {author} {\bibinfo {author} {\bibfnamefont {B.~I.}\ \bibnamefont
  {Abelev}} \emph {et~al.} (\bibinfo {collaboration} {STAR}),\ }\href {\doibase
  10.1103/PhysRevC.77.061902} {\bibfield  {journal} {\bibinfo  {journal} {Phys.
  Rev. C}\ }\textbf {\bibinfo {volume} {77}},\ \bibinfo {pages} {061902}
  (\bibinfo {year} {2008})},\ \Eprint {http://arxiv.org/abs/0801.1729}
  {arXiv:0801.1729 [nucl-ex]} \BibitemShut {NoStop}%
\bibitem [{\citenamefont {Acharya}\ \emph {et~al.}(2020)\citenamefont {Acharya}
  \emph {et~al.}}]{ALICE:2019aid}%
  \BibitemOpen
  \bibfield  {author} {\bibinfo {author} {\bibfnamefont {S.}~\bibnamefont
  {Acharya}} \emph {et~al.} (\bibinfo {collaboration} {ALICE}),\ }\href
  {\doibase 10.1103/PhysRevLett.125.012301} {\bibfield  {journal} {\bibinfo
  {journal} {Phys. Rev. Lett.}\ }\textbf {\bibinfo {volume} {125}},\ \bibinfo
  {pages} {012301} (\bibinfo {year} {2020})},\ \Eprint
  {http://arxiv.org/abs/1910.14408} {arXiv:1910.14408 [nucl-ex]} \BibitemShut
  {NoStop}%
\bibitem [{\citenamefont {Abdallah}\ \emph {et~al.}(2023)\citenamefont
  {Abdallah} \emph {et~al.}}]{STAR:2022fan}%
  \BibitemOpen
  \bibfield  {author} {\bibinfo {author} {\bibfnamefont {M.~S.}\ \bibnamefont
  {Abdallah}} \emph {et~al.} (\bibinfo {collaboration} {STAR}),\ }\href
  {\doibase 10.1038/s41586-022-05557-5} {\bibfield  {journal} {\bibinfo
  {journal} {Nature}\ }\textbf {\bibinfo {volume} {614}},\ \bibinfo {pages}
  {244} (\bibinfo {year} {2023})},\ \Eprint {http://arxiv.org/abs/2204.02302}
  {arXiv:2204.02302 [hep-ph]} \BibitemShut {NoStop}%
\bibitem [{\citenamefont {Acharya}\ \emph {et~al.}(2023)\citenamefont {Acharya}
  \emph {et~al.}}]{ALICE:2022dyy}%
  \BibitemOpen
  \bibfield  {author} {\bibinfo {author} {\bibfnamefont {S.}~\bibnamefont
  {Acharya}} \emph {et~al.} (\bibinfo {collaboration} {ALICE}),\ }\href
  {\doibase 10.1103/PhysRevLett.131.042303} {\bibfield  {journal} {\bibinfo
  {journal} {Phys. Rev. Lett.}\ }\textbf {\bibinfo {volume} {131}},\ \bibinfo
  {pages} {042303} (\bibinfo {year} {2023})},\ \Eprint
  {http://arxiv.org/abs/2204.10171} {arXiv:2204.10171 [nucl-ex]} \BibitemShut
  {NoStop}%
\bibitem [{\citenamefont {Becattini}\ and\ \citenamefont
  {Piccinini}(2008)}]{Becattini:2007nd}%
  \BibitemOpen
  \bibfield  {author} {\bibinfo {author} {\bibfnamefont {F.}~\bibnamefont
  {Becattini}}\ and\ \bibinfo {author} {\bibfnamefont {F.}~\bibnamefont
  {Piccinini}},\ }\href {\doibase 10.1016/j.aop.2008.01.001} {\bibfield
  {journal} {\bibinfo  {journal} {Annals Phys.}\ }\textbf {\bibinfo {volume}
  {323}},\ \bibinfo {pages} {2452} (\bibinfo {year} {2008})},\ \Eprint
  {http://arxiv.org/abs/0710.5694} {arXiv:0710.5694 [nucl-th]} \BibitemShut
  {NoStop}%
\bibitem [{\citenamefont {Becattini}\ \emph {et~al.}(2008)\citenamefont
  {Becattini}, \citenamefont {Piccinini},\ and\ \citenamefont
  {Rizzo}}]{Becattini:2007sr}%
  \BibitemOpen
  \bibfield  {author} {\bibinfo {author} {\bibfnamefont {F.}~\bibnamefont
  {Becattini}}, \bibinfo {author} {\bibfnamefont {F.}~\bibnamefont
  {Piccinini}}, \ and\ \bibinfo {author} {\bibfnamefont {J.}~\bibnamefont
  {Rizzo}},\ }\href {\doibase 10.1103/PhysRevC.77.024906} {\bibfield  {journal}
  {\bibinfo  {journal} {Phys. Rev. C}\ }\textbf {\bibinfo {volume} {77}},\
  \bibinfo {pages} {024906} (\bibinfo {year} {2008})},\ \Eprint
  {http://arxiv.org/abs/0711.1253} {arXiv:0711.1253 [nucl-th]} \BibitemShut
  {NoStop}%
\bibitem [{\citenamefont {Xia}\ \emph {et~al.}(2021)\citenamefont {Xia},
  \citenamefont {Li}, \citenamefont {Huang},\ and\ \citenamefont
  {Zhong~Huang}}]{Xia:2020tyd}%
  \BibitemOpen
  \bibfield  {author} {\bibinfo {author} {\bibfnamefont {X.-L.}\ \bibnamefont
  {Xia}}, \bibinfo {author} {\bibfnamefont {H.}~\bibnamefont {Li}}, \bibinfo
  {author} {\bibfnamefont {X.-G.}\ \bibnamefont {Huang}}, \ and\ \bibinfo
  {author} {\bibfnamefont {H.}~\bibnamefont {Zhong~Huang}},\ }\href {\doibase
  10.1016/j.physletb.2021.136325} {\bibfield  {journal} {\bibinfo  {journal}
  {Phys. Lett. B}\ }\textbf {\bibinfo {volume} {817}},\ \bibinfo {pages}
  {136325} (\bibinfo {year} {2021})},\ \Eprint
  {http://arxiv.org/abs/2010.01474} {arXiv:2010.01474 [nucl-th]} \BibitemShut
  {NoStop}%
\bibitem [{\citenamefont {Becattini}(2022)}]{Becattini:2022zvf}%
  \BibitemOpen
  \bibfield  {author} {\bibinfo {author} {\bibfnamefont {F.}~\bibnamefont
  {Becattini}},\ }\href {\doibase 10.1088/1361-6633/ac97a9} {\bibfield
  {journal} {\bibinfo  {journal} {Rept. Prog. Phys.}\ }\textbf {\bibinfo
  {volume} {85}},\ \bibinfo {pages} {122301} (\bibinfo {year} {2022})},\
  \Eprint {http://arxiv.org/abs/2204.01144} {arXiv:2204.01144 [nucl-th]}
  \BibitemShut {NoStop}%
\bibitem [{\citenamefont {Sheng}\ \emph {et~al.}(2020)\citenamefont {Sheng},
  \citenamefont {Oliva},\ and\ \citenamefont {Wang}}]{Sheng:2019kmk}%
  \BibitemOpen
  \bibfield  {author} {\bibinfo {author} {\bibfnamefont {X.-L.}\ \bibnamefont
  {Sheng}}, \bibinfo {author} {\bibfnamefont {L.}~\bibnamefont {Oliva}}, \ and\
  \bibinfo {author} {\bibfnamefont {Q.}~\bibnamefont {Wang}},\ }\href {\doibase
  10.1103/PhysRevD.101.096005} {\bibfield  {journal} {\bibinfo  {journal}
  {Phys. Rev. D}\ }\textbf {\bibinfo {volume} {101}},\ \bibinfo {pages}
  {096005} (\bibinfo {year} {2020})},\ \Eprint
  {http://arxiv.org/abs/1910.13684} {arXiv:1910.13684 [nucl-th]} \BibitemShut
  {NoStop}%
\bibitem [{\citenamefont {M\"uller}\ and\ \citenamefont
  {Yang}(2022)}]{Muller:2021hpe}%
  \BibitemOpen
  \bibfield  {author} {\bibinfo {author} {\bibfnamefont {B.}~\bibnamefont
  {M\"uller}}\ and\ \bibinfo {author} {\bibfnamefont {D.-L.}\ \bibnamefont
  {Yang}},\ }\href {\doibase 10.1103/PhysRevD.105.L011901} {\bibfield
  {journal} {\bibinfo  {journal} {Phys. Rev. D}\ }\textbf {\bibinfo {volume}
  {105}},\ \bibinfo {pages} {L011901} (\bibinfo {year} {2022})},\ \bibinfo
  {note} {[Erratum: Phys.Rev.D 106, 039904 (2022)]},\ \Eprint
  {http://arxiv.org/abs/2110.15630} {arXiv:2110.15630 [nucl-th]} \BibitemShut
  {NoStop}%
\bibitem [{\citenamefont {Sheng}\ \emph {et~al.}(2022)\citenamefont {Sheng},
  \citenamefont {Oliva}, \citenamefont {Liang}, \citenamefont {Wang},\ and\
  \citenamefont {Wang}}]{Sheng:2022ffb}%
  \BibitemOpen
  \bibfield  {author} {\bibinfo {author} {\bibfnamefont {X.-L.}\ \bibnamefont
  {Sheng}}, \bibinfo {author} {\bibfnamefont {L.}~\bibnamefont {Oliva}},
  \bibinfo {author} {\bibfnamefont {Z.-T.}\ \bibnamefont {Liang}}, \bibinfo
  {author} {\bibfnamefont {Q.}~\bibnamefont {Wang}}, \ and\ \bibinfo {author}
  {\bibfnamefont {X.-N.}\ \bibnamefont {Wang}},\ }\href@noop {} {\  (\bibinfo
  {year} {2022})},\ \Eprint {http://arxiv.org/abs/2206.05868} {arXiv:2206.05868
  [hep-ph]} \BibitemShut {NoStop}%
\bibitem [{\citenamefont {Sheng}\ \emph {et~al.}(2023)\citenamefont {Sheng},
  \citenamefont {Oliva}, \citenamefont {Liang}, \citenamefont {Wang},\ and\
  \citenamefont {Wang}}]{Sheng:2022wsy}%
  \BibitemOpen
  \bibfield  {author} {\bibinfo {author} {\bibfnamefont {X.-L.}\ \bibnamefont
  {Sheng}}, \bibinfo {author} {\bibfnamefont {L.}~\bibnamefont {Oliva}},
  \bibinfo {author} {\bibfnamefont {Z.-T.}\ \bibnamefont {Liang}}, \bibinfo
  {author} {\bibfnamefont {Q.}~\bibnamefont {Wang}}, \ and\ \bibinfo {author}
  {\bibfnamefont {X.-N.}\ \bibnamefont {Wang}},\ }\href {\doibase
  10.1103/PhysRevLett.131.042304} {\bibfield  {journal} {\bibinfo  {journal}
  {Phys. Rev. Lett.}\ }\textbf {\bibinfo {volume} {131}},\ \bibinfo {pages}
  {042304} (\bibinfo {year} {2023})},\ \Eprint
  {http://arxiv.org/abs/2205.15689} {arXiv:2205.15689 [nucl-th]} \BibitemShut
  {NoStop}%
\bibitem [{\citenamefont {Li}\ and\ \citenamefont {Liu}(2022)}]{Li:2022vmb}%
  \BibitemOpen
  \bibfield  {author} {\bibinfo {author} {\bibfnamefont {F.}~\bibnamefont
  {Li}}\ and\ \bibinfo {author} {\bibfnamefont {S.~Y.~F.}\ \bibnamefont
  {Liu}},\ }\href@noop {} {\  (\bibinfo {year} {2022})},\ \Eprint
  {http://arxiv.org/abs/2206.11890} {arXiv:2206.11890 [nucl-th]} \BibitemShut
  {NoStop}%
\bibitem [{\citenamefont {Wagner}\ \emph {et~al.}(2023)\citenamefont {Wagner},
  \citenamefont {Weickgenannt},\ and\ \citenamefont
  {Speranza}}]{Wagner:2022gza}%
  \BibitemOpen
  \bibfield  {author} {\bibinfo {author} {\bibfnamefont {D.}~\bibnamefont
  {Wagner}}, \bibinfo {author} {\bibfnamefont {N.}~\bibnamefont
  {Weickgenannt}}, \ and\ \bibinfo {author} {\bibfnamefont {E.}~\bibnamefont
  {Speranza}},\ }\href {\doibase 10.1103/PhysRevResearch.5.013187} {\bibfield
  {journal} {\bibinfo  {journal} {Phys. Rev. Res.}\ }\textbf {\bibinfo {volume}
  {5}},\ \bibinfo {pages} {013187} (\bibinfo {year} {2023})},\ \Eprint
  {http://arxiv.org/abs/2207.01111} {arXiv:2207.01111 [nucl-th]} \BibitemShut
  {NoStop}%
\bibitem [{\citenamefont {Wei}\ and\ \citenamefont
  {Huang}(2023)}]{Wei:2023pdf}%
  \BibitemOpen
  \bibfield  {author} {\bibinfo {author} {\bibfnamefont {M.}~\bibnamefont
  {Wei}}\ and\ \bibinfo {author} {\bibfnamefont {M.}~\bibnamefont {Huang}},\
  }\href {\doibase 10.1088/1674-1137/acf036} {\bibfield  {journal} {\bibinfo
  {journal} {Chin. Phys. C}\ }\textbf {\bibinfo {volume} {47}},\ \bibinfo
  {pages} {104105} (\bibinfo {year} {2023})},\ \Eprint
  {http://arxiv.org/abs/2303.01897} {arXiv:2303.01897 [hep-ph]} \BibitemShut
  {NoStop}%
\bibitem [{\citenamefont {Dong}\ \emph {et~al.}(2024)\citenamefont {Dong},
  \citenamefont {Yin}, \citenamefont {Sheng}, \citenamefont {Yang},\ and\
  \citenamefont {Wang}}]{Dong:2023cng}%
  \BibitemOpen
  \bibfield  {author} {\bibinfo {author} {\bibfnamefont {W.-B.}\ \bibnamefont
  {Dong}}, \bibinfo {author} {\bibfnamefont {Y.-L.}\ \bibnamefont {Yin}},
  \bibinfo {author} {\bibfnamefont {X.-L.}\ \bibnamefont {Sheng}}, \bibinfo
  {author} {\bibfnamefont {S.-Z.}\ \bibnamefont {Yang}}, \ and\ \bibinfo
  {author} {\bibfnamefont {Q.}~\bibnamefont {Wang}},\ }\href {\doibase
  10.1103/PhysRevD.109.056025} {\bibfield  {journal} {\bibinfo  {journal}
  {Phys. Rev. D}\ }\textbf {\bibinfo {volume} {109}},\ \bibinfo {pages}
  {056025} (\bibinfo {year} {2024})},\ \Eprint
  {http://arxiv.org/abs/2311.18400} {arXiv:2311.18400 [hep-ph]} \BibitemShut
  {NoStop}%
\bibitem [{\citenamefont {Yin}\ \emph {et~al.}(2025)\citenamefont {Yin},
  \citenamefont {Dong}, \citenamefont {Yi},\ and\ \citenamefont
  {Wang}}]{Yin:2025gvl}%
  \BibitemOpen
  \bibfield  {author} {\bibinfo {author} {\bibfnamefont {Y.-L.}\ \bibnamefont
  {Yin}}, \bibinfo {author} {\bibfnamefont {W.-B.}\ \bibnamefont {Dong}},
  \bibinfo {author} {\bibfnamefont {C.}~\bibnamefont {Yi}}, \ and\ \bibinfo
  {author} {\bibfnamefont {Q.}~\bibnamefont {Wang}},\ }\href@noop {} {\
  (\bibinfo {year} {2025})},\ \Eprint {http://arxiv.org/abs/2503.09937}
  {arXiv:2503.09937 [hep-ph]} \BibitemShut {NoStop}%
\bibitem [{\citenamefont {Li}\ and\ \citenamefont {Liu}(2025)}]{Li:2025pef}%
  \BibitemOpen
  \bibfield  {author} {\bibinfo {author} {\bibfnamefont {Y.}~\bibnamefont
  {Li}}\ and\ \bibinfo {author} {\bibfnamefont {S.~Y.~F.}\ \bibnamefont
  {Liu}},\ }\href@noop {} {\  (\bibinfo {year} {2025})},\ \Eprint
  {http://arxiv.org/abs/2501.17861} {arXiv:2501.17861 [nucl-th]} \BibitemShut
  {NoStop}%
\bibitem [{\citenamefont {Agakichiev}\ \emph {et~al.}(1995)\citenamefont
  {Agakichiev} \emph {et~al.}}]{CERES:1995vll}%
  \BibitemOpen
  \bibfield  {author} {\bibinfo {author} {\bibfnamefont {G.}~\bibnamefont
  {Agakichiev}} \emph {et~al.} (\bibinfo {collaboration} {CERES}),\ }\href
  {\doibase 10.1103/PhysRevLett.75.1272} {\bibfield  {journal} {\bibinfo
  {journal} {Phys. Rev. Lett.}\ }\textbf {\bibinfo {volume} {75}},\ \bibinfo
  {pages} {1272} (\bibinfo {year} {1995})}\BibitemShut {NoStop}%
\bibitem [{\citenamefont {Li}\ \emph {et~al.}(1995)\citenamefont {Li},
  \citenamefont {Ko},\ and\ \citenamefont {Brown}}]{Li:1995qm}%
  \BibitemOpen
  \bibfield  {author} {\bibinfo {author} {\bibfnamefont {G.-Q.}\ \bibnamefont
  {Li}}, \bibinfo {author} {\bibfnamefont {C.~M.}\ \bibnamefont {Ko}}, \ and\
  \bibinfo {author} {\bibfnamefont {G.~E.}\ \bibnamefont {Brown}},\ }\href
  {\doibase 10.1103/PhysRevLett.75.4007} {\bibfield  {journal} {\bibinfo
  {journal} {Phys. Rev. Lett.}\ }\textbf {\bibinfo {volume} {75}},\ \bibinfo
  {pages} {4007} (\bibinfo {year} {1995})},\ \Eprint
  {http://arxiv.org/abs/nucl-th/9504025} {arXiv:nucl-th/9504025} \BibitemShut
  {NoStop}%
\bibitem [{\citenamefont {Cassing}\ \emph {et~al.}(1995)\citenamefont
  {Cassing}, \citenamefont {Ehehalt},\ and\ \citenamefont
  {Ko}}]{Cassing:1995zv}%
  \BibitemOpen
  \bibfield  {author} {\bibinfo {author} {\bibfnamefont {W.}~\bibnamefont
  {Cassing}}, \bibinfo {author} {\bibfnamefont {W.}~\bibnamefont {Ehehalt}}, \
  and\ \bibinfo {author} {\bibfnamefont {C.~M.}\ \bibnamefont {Ko}},\ }\href
  {\doibase 10.1016/0370-2693(95)01214-B} {\bibfield  {journal} {\bibinfo
  {journal} {Phys. Lett. B}\ }\textbf {\bibinfo {volume} {363}},\ \bibinfo
  {pages} {35} (\bibinfo {year} {1995})},\ \Eprint
  {http://arxiv.org/abs/hep-ph/9508233} {arXiv:hep-ph/9508233} \BibitemShut
  {NoStop}%
\bibitem [{\citenamefont {Hatsuda}\ and\ \citenamefont
  {Lee}(1992)}]{Hatsuda:1991ez}%
  \BibitemOpen
  \bibfield  {author} {\bibinfo {author} {\bibfnamefont {T.}~\bibnamefont
  {Hatsuda}}\ and\ \bibinfo {author} {\bibfnamefont {S.~H.}\ \bibnamefont
  {Lee}},\ }\href {\doibase 10.1103/PhysRevC.46.R34} {\bibfield  {journal}
  {\bibinfo  {journal} {Phys. Rev. C}\ }\textbf {\bibinfo {volume} {46}},\
  \bibinfo {pages} {R34} (\bibinfo {year} {1992})}\BibitemShut {NoStop}%
\bibitem [{\citenamefont {Hatsuda}\ \emph {et~al.}(1993)\citenamefont
  {Hatsuda}, \citenamefont {Koike},\ and\ \citenamefont
  {Lee}}]{Hatsuda:1992bv}%
  \BibitemOpen
  \bibfield  {author} {\bibinfo {author} {\bibfnamefont {T.}~\bibnamefont
  {Hatsuda}}, \bibinfo {author} {\bibfnamefont {Y.}~\bibnamefont {Koike}}, \
  and\ \bibinfo {author} {\bibfnamefont {S.-H.}\ \bibnamefont {Lee}},\ }\href
  {\doibase 10.1016/0550-3213(93)90107-Z} {\bibfield  {journal} {\bibinfo
  {journal} {Nucl. Phys. B}\ }\textbf {\bibinfo {volume} {394}},\ \bibinfo
  {pages} {221} (\bibinfo {year} {1993})}\BibitemShut {NoStop}%
\bibitem [{\citenamefont {Brown}\ and\ \citenamefont
  {Rho}(1991)}]{Brown:1991kk}%
  \BibitemOpen
  \bibfield  {author} {\bibinfo {author} {\bibfnamefont {G.~E.}\ \bibnamefont
  {Brown}}\ and\ \bibinfo {author} {\bibfnamefont {M.}~\bibnamefont {Rho}},\
  }\href {\doibase 10.1103/PhysRevLett.66.2720} {\bibfield  {journal} {\bibinfo
   {journal} {Phys. Rev. Lett.}\ }\textbf {\bibinfo {volume} {66}},\ \bibinfo
  {pages} {2720} (\bibinfo {year} {1991})}\BibitemShut {NoStop}%
\bibitem [{\citenamefont {Chanfray}\ \emph {et~al.}(1996)\citenamefont
  {Chanfray}, \citenamefont {Rapp},\ and\ \citenamefont
  {Wambach}}]{Chanfray:1995jgo}%
  \BibitemOpen
  \bibfield  {author} {\bibinfo {author} {\bibfnamefont {G.}~\bibnamefont
  {Chanfray}}, \bibinfo {author} {\bibfnamefont {R.}~\bibnamefont {Rapp}}, \
  and\ \bibinfo {author} {\bibfnamefont {J.}~\bibnamefont {Wambach}},\ }\href
  {\doibase 10.1103/PhysRevLett.76.368} {\bibfield  {journal} {\bibinfo
  {journal} {Phys. Rev. Lett.}\ }\textbf {\bibinfo {volume} {76}},\ \bibinfo
  {pages} {368} (\bibinfo {year} {1996})},\ \Eprint
  {http://arxiv.org/abs/hep-ph/9508353} {arXiv:hep-ph/9508353} \BibitemShut
  {NoStop}%
\bibitem [{\citenamefont {Rapp}\ \emph {et~al.}(1997)\citenamefont {Rapp},
  \citenamefont {Chanfray},\ and\ \citenamefont {Wambach}}]{Rapp:1997fs}%
  \BibitemOpen
  \bibfield  {author} {\bibinfo {author} {\bibfnamefont {R.}~\bibnamefont
  {Rapp}}, \bibinfo {author} {\bibfnamefont {G.}~\bibnamefont {Chanfray}}, \
  and\ \bibinfo {author} {\bibfnamefont {J.}~\bibnamefont {Wambach}},\ }\href
  {\doibase 10.1016/S0375-9474(97)00137-1} {\bibfield  {journal} {\bibinfo
  {journal} {Nucl. Phys. A}\ }\textbf {\bibinfo {volume} {617}},\ \bibinfo
  {pages} {472} (\bibinfo {year} {1997})},\ \Eprint
  {http://arxiv.org/abs/hep-ph/9702210} {arXiv:hep-ph/9702210} \BibitemShut
  {NoStop}%
\bibitem [{\citenamefont {Arnaldi}\ \emph {et~al.}(2006)\citenamefont {Arnaldi}
  \emph {et~al.}}]{NA60:2006ymb}%
  \BibitemOpen
  \bibfield  {author} {\bibinfo {author} {\bibfnamefont {R.}~\bibnamefont
  {Arnaldi}} \emph {et~al.} (\bibinfo {collaboration} {NA60}),\ }\href
  {\doibase 10.1103/PhysRevLett.96.162302} {\bibfield  {journal} {\bibinfo
  {journal} {Phys. Rev. Lett.}\ }\textbf {\bibinfo {volume} {96}},\ \bibinfo
  {pages} {162302} (\bibinfo {year} {2006})},\ \Eprint
  {http://arxiv.org/abs/nucl-ex/0605007} {arXiv:nucl-ex/0605007} \BibitemShut
  {NoStop}%
\bibitem [{\citenamefont {Adare}\ \emph {et~al.}(2010)\citenamefont {Adare}
  \emph {et~al.}}]{PHENIX:2009gyd}%
  \BibitemOpen
  \bibfield  {author} {\bibinfo {author} {\bibfnamefont {A.}~\bibnamefont
  {Adare}} \emph {et~al.} (\bibinfo {collaboration} {PHENIX}),\ }\href
  {\doibase 10.1103/PhysRevC.81.034911} {\bibfield  {journal} {\bibinfo
  {journal} {Phys. Rev. C}\ }\textbf {\bibinfo {volume} {81}},\ \bibinfo
  {pages} {034911} (\bibinfo {year} {2010})},\ \Eprint
  {http://arxiv.org/abs/0912.0244} {arXiv:0912.0244 [nucl-ex]} \BibitemShut
  {NoStop}%
\bibitem [{\citenamefont {Adamczyk}\ \emph {et~al.}(2014)\citenamefont
  {Adamczyk} \emph {et~al.}}]{STAR:2013pwb}%
  \BibitemOpen
  \bibfield  {author} {\bibinfo {author} {\bibfnamefont {L.}~\bibnamefont
  {Adamczyk}} \emph {et~al.} (\bibinfo {collaboration} {STAR}),\ }\href
  {\doibase 10.1103/PhysRevLett.113.022301} {\bibfield  {journal} {\bibinfo
  {journal} {Phys. Rev. Lett.}\ }\textbf {\bibinfo {volume} {113}},\ \bibinfo
  {pages} {022301} (\bibinfo {year} {2014})},\ \bibinfo {note} {[Addendum:
  Phys.Rev.Lett. 113, 049903 (2014)]},\ \Eprint
  {http://arxiv.org/abs/1312.7397} {arXiv:1312.7397 [hep-ex]} \BibitemShut
  {NoStop}%
\bibitem [{\citenamefont {Adare}\ \emph {et~al.}(2016)\citenamefont {Adare}
  \emph {et~al.}}]{PHENIX:2015vek}%
  \BibitemOpen
  \bibfield  {author} {\bibinfo {author} {\bibfnamefont {A.}~\bibnamefont
  {Adare}} \emph {et~al.} (\bibinfo {collaboration} {PHENIX}),\ }\href
  {\doibase 10.1103/PhysRevC.93.014904} {\bibfield  {journal} {\bibinfo
  {journal} {Phys. Rev. C}\ }\textbf {\bibinfo {volume} {93}},\ \bibinfo
  {pages} {014904} (\bibinfo {year} {2016})},\ \Eprint
  {http://arxiv.org/abs/1509.04667} {arXiv:1509.04667 [nucl-ex]} \BibitemShut
  {NoStop}%
\bibitem [{\citenamefont {Acharya}\ \emph {et~al.}(2019)\citenamefont {Acharya}
  \emph {et~al.}}]{ALICE:2018ael}%
  \BibitemOpen
  \bibfield  {author} {\bibinfo {author} {\bibfnamefont {S.}~\bibnamefont
  {Acharya}} \emph {et~al.} (\bibinfo {collaboration} {ALICE}),\ }\href
  {\doibase 10.1103/PhysRevC.99.024002} {\bibfield  {journal} {\bibinfo
  {journal} {Phys. Rev. C}\ }\textbf {\bibinfo {volume} {99}},\ \bibinfo
  {pages} {024002} (\bibinfo {year} {2019})},\ \Eprint
  {http://arxiv.org/abs/1807.00923} {arXiv:1807.00923 [nucl-ex]} \BibitemShut
  {NoStop}%
\bibitem [{\citenamefont {Rapp}(2001)}]{Rapp:2000pe}%
  \BibitemOpen
  \bibfield  {author} {\bibinfo {author} {\bibfnamefont {R.}~\bibnamefont
  {Rapp}},\ }\href {\doibase 10.1103/PhysRevC.63.054907} {\bibfield  {journal}
  {\bibinfo  {journal} {Phys. Rev. C}\ }\textbf {\bibinfo {volume} {63}},\
  \bibinfo {pages} {054907} (\bibinfo {year} {2001})},\ \Eprint
  {http://arxiv.org/abs/hep-ph/0010101} {arXiv:hep-ph/0010101} \BibitemShut
  {NoStop}%
\bibitem [{\citenamefont {van Hees}\ and\ \citenamefont
  {Rapp}(2006)}]{vanHees:2006ng}%
  \BibitemOpen
  \bibfield  {author} {\bibinfo {author} {\bibfnamefont {H.}~\bibnamefont {van
  Hees}}\ and\ \bibinfo {author} {\bibfnamefont {R.}~\bibnamefont {Rapp}},\
  }\href {\doibase 10.1103/PhysRevLett.97.102301} {\bibfield  {journal}
  {\bibinfo  {journal} {Phys. Rev. Lett.}\ }\textbf {\bibinfo {volume} {97}},\
  \bibinfo {pages} {102301} (\bibinfo {year} {2006})},\ \Eprint
  {http://arxiv.org/abs/hep-ph/0603084} {arXiv:hep-ph/0603084} \BibitemShut
  {NoStop}%
\bibitem [{\citenamefont {van Hees}\ and\ \citenamefont
  {Rapp}(2008)}]{vanHees:2007th}%
  \BibitemOpen
  \bibfield  {author} {\bibinfo {author} {\bibfnamefont {H.}~\bibnamefont {van
  Hees}}\ and\ \bibinfo {author} {\bibfnamefont {R.}~\bibnamefont {Rapp}},\
  }\href {\doibase 10.1016/j.nuclphysa.2008.03.009} {\bibfield  {journal}
  {\bibinfo  {journal} {Nucl. Phys. A}\ }\textbf {\bibinfo {volume} {806}},\
  \bibinfo {pages} {339} (\bibinfo {year} {2008})},\ \Eprint
  {http://arxiv.org/abs/0711.3444} {arXiv:0711.3444 [hep-ph]} \BibitemShut
  {NoStop}%
\bibitem [{\citenamefont {Rapp}(2013)}]{Rapp:2013ema}%
  \BibitemOpen
  \bibfield  {author} {\bibinfo {author} {\bibfnamefont {R.}~\bibnamefont
  {Rapp}},\ }\href {\doibase 10.22323/1.185.0008} {\bibfield  {journal}
  {\bibinfo  {journal} {PoS}\ }\textbf {\bibinfo {volume} {CPOD2013}},\
  \bibinfo {pages} {008} (\bibinfo {year} {2013})},\ \Eprint
  {http://arxiv.org/abs/1306.6394} {arXiv:1306.6394 [nucl-th]} \BibitemShut
  {NoStop}%
\bibitem [{\citenamefont {Linnyk}\ \emph {et~al.}(2011)\citenamefont {Linnyk},
  \citenamefont {Bratkovskaya}, \citenamefont {Ozvenchuk}, \citenamefont
  {Cassing},\ and\ \citenamefont {Ko}}]{Linnyk:2011hz}%
  \BibitemOpen
  \bibfield  {author} {\bibinfo {author} {\bibfnamefont {O.}~\bibnamefont
  {Linnyk}}, \bibinfo {author} {\bibfnamefont {E.~L.}\ \bibnamefont
  {Bratkovskaya}}, \bibinfo {author} {\bibfnamefont {V.}~\bibnamefont
  {Ozvenchuk}}, \bibinfo {author} {\bibfnamefont {W.}~\bibnamefont {Cassing}},
  \ and\ \bibinfo {author} {\bibfnamefont {C.~M.}\ \bibnamefont {Ko}},\ }\href
  {\doibase 10.1103/PhysRevC.84.054917} {\bibfield  {journal} {\bibinfo
  {journal} {Phys. Rev. C}\ }\textbf {\bibinfo {volume} {84}},\ \bibinfo
  {pages} {054917} (\bibinfo {year} {2011})},\ \Eprint
  {http://arxiv.org/abs/1107.3402} {arXiv:1107.3402 [nucl-th]} \BibitemShut
  {NoStop}%
\bibitem [{\citenamefont {Linnyk}\ \emph {et~al.}(2012)\citenamefont {Linnyk},
  \citenamefont {Cassing}, \citenamefont {Manninen}, \citenamefont
  {Bratkovskaya},\ and\ \citenamefont {Ko}}]{Linnyk:2011vx}%
  \BibitemOpen
  \bibfield  {author} {\bibinfo {author} {\bibfnamefont {O.}~\bibnamefont
  {Linnyk}}, \bibinfo {author} {\bibfnamefont {W.}~\bibnamefont {Cassing}},
  \bibinfo {author} {\bibfnamefont {J.}~\bibnamefont {Manninen}}, \bibinfo
  {author} {\bibfnamefont {E.~L.}\ \bibnamefont {Bratkovskaya}}, \ and\
  \bibinfo {author} {\bibfnamefont {C.~M.}\ \bibnamefont {Ko}},\ }\href
  {\doibase 10.1103/PhysRevC.85.024910} {\bibfield  {journal} {\bibinfo
  {journal} {Phys. Rev. C}\ }\textbf {\bibinfo {volume} {85}},\ \bibinfo
  {pages} {024910} (\bibinfo {year} {2012})},\ \Eprint
  {http://arxiv.org/abs/1111.2975} {arXiv:1111.2975 [nucl-th]} \BibitemShut
  {NoStop}%
\bibitem [{\citenamefont {Cassing}\ and\ \citenamefont
  {Bratkovskaya}(2009)}]{Cassing:2009vt}%
  \BibitemOpen
  \bibfield  {author} {\bibinfo {author} {\bibfnamefont {W.}~\bibnamefont
  {Cassing}}\ and\ \bibinfo {author} {\bibfnamefont {E.~L.}\ \bibnamefont
  {Bratkovskaya}},\ }\href {\doibase 10.1016/j.nuclphysa.2009.09.007}
  {\bibfield  {journal} {\bibinfo  {journal} {Nucl. Phys. A}\ }\textbf
  {\bibinfo {volume} {831}},\ \bibinfo {pages} {215} (\bibinfo {year}
  {2009})},\ \Eprint {http://arxiv.org/abs/0907.5331} {arXiv:0907.5331
  [nucl-th]} \BibitemShut {NoStop}%
\bibitem [{\citenamefont {Bratkovskaya}\ \emph {et~al.}(2011)\citenamefont
  {Bratkovskaya}, \citenamefont {Cassing}, \citenamefont {Konchakovski},\ and\
  \citenamefont {Linnyk}}]{Bratkovskaya:2011wp}%
  \BibitemOpen
  \bibfield  {author} {\bibinfo {author} {\bibfnamefont {E.~L.}\ \bibnamefont
  {Bratkovskaya}}, \bibinfo {author} {\bibfnamefont {W.}~\bibnamefont
  {Cassing}}, \bibinfo {author} {\bibfnamefont {V.~P.}\ \bibnamefont
  {Konchakovski}}, \ and\ \bibinfo {author} {\bibfnamefont {O.}~\bibnamefont
  {Linnyk}},\ }\href {\doibase 10.1016/j.nuclphysa.2011.03.003} {\bibfield
  {journal} {\bibinfo  {journal} {Nucl. Phys. A}\ }\textbf {\bibinfo {volume}
  {856}},\ \bibinfo {pages} {162} (\bibinfo {year} {2011})},\ \Eprint
  {http://arxiv.org/abs/1101.5793} {arXiv:1101.5793 [nucl-th]} \BibitemShut
  {NoStop}%
\bibitem [{\citenamefont {Alberico}\ \emph {et~al.}(2007)\citenamefont
  {Alberico}, \citenamefont {Beraudo}, \citenamefont {De~Pace},\ and\
  \citenamefont {Molinari}}]{Alberico:2006vw}%
  \BibitemOpen
  \bibfield  {author} {\bibinfo {author} {\bibfnamefont {W.~M.}\ \bibnamefont
  {Alberico}}, \bibinfo {author} {\bibfnamefont {A.}~\bibnamefont {Beraudo}},
  \bibinfo {author} {\bibfnamefont {A.}~\bibnamefont {De~Pace}}, \ and\
  \bibinfo {author} {\bibfnamefont {A.}~\bibnamefont {Molinari}},\ }\href
  {\doibase 10.1103/PhysRevD.75.074009} {\bibfield  {journal} {\bibinfo
  {journal} {Phys. Rev. D}\ }\textbf {\bibinfo {volume} {75}},\ \bibinfo
  {pages} {074009} (\bibinfo {year} {2007})}\BibitemShut {NoStop}%
\bibitem [{\citenamefont {Aarts}\ \emph {et~al.}(2011)\citenamefont {Aarts},
  \citenamefont {Allton}, \citenamefont {Kim}, \citenamefont {Lombardo},
  \citenamefont {Oktay}, \citenamefont {Ryan}, \citenamefont {Sinclair},\ and\
  \citenamefont {Skullerud}}]{Aarts:2011sm}%
  \BibitemOpen
  \bibfield  {author} {\bibinfo {author} {\bibfnamefont {G.}~\bibnamefont
  {Aarts}}, \bibinfo {author} {\bibfnamefont {C.}~\bibnamefont {Allton}},
  \bibinfo {author} {\bibfnamefont {S.}~\bibnamefont {Kim}}, \bibinfo {author}
  {\bibfnamefont {M.~P.}\ \bibnamefont {Lombardo}}, \bibinfo {author}
  {\bibfnamefont {M.~B.}\ \bibnamefont {Oktay}}, \bibinfo {author}
  {\bibfnamefont {S.~M.}\ \bibnamefont {Ryan}}, \bibinfo {author}
  {\bibfnamefont {D.~K.}\ \bibnamefont {Sinclair}}, \ and\ \bibinfo {author}
  {\bibfnamefont {J.~I.}\ \bibnamefont {Skullerud}},\ }\href@noop {} {\bibfield
   {journal} {\bibinfo  {journal} {JHEP}\ }\textbf {\bibinfo {volume} {11}},\
  \bibinfo {pages} {103} (\bibinfo {year} {2011})}\BibitemShut {NoStop}%
\bibitem [{\citenamefont {Ding}\ \emph {et~al.}(2012)\citenamefont {Ding},
  \citenamefont {Francis}, \citenamefont {Kaczmarek}, \citenamefont {Karsch},
  \citenamefont {Satz},\ and\ \citenamefont {Soeldner}}]{Ding:2012sp}%
  \BibitemOpen
  \bibfield  {author} {\bibinfo {author} {\bibfnamefont {H.~T.}\ \bibnamefont
  {Ding}}, \bibinfo {author} {\bibfnamefont {A.}~\bibnamefont {Francis}},
  \bibinfo {author} {\bibfnamefont {O.}~\bibnamefont {Kaczmarek}}, \bibinfo
  {author} {\bibfnamefont {F.}~\bibnamefont {Karsch}}, \bibinfo {author}
  {\bibfnamefont {H.}~\bibnamefont {Satz}}, \ and\ \bibinfo {author}
  {\bibfnamefont {W.}~\bibnamefont {Soeldner}},\ }\href@noop {} {\bibfield
  {journal} {\bibinfo  {journal} {Phys. Rev. D}\ }\textbf {\bibinfo {volume}
  {86}},\ \bibinfo {pages} {014509} (\bibinfo {year} {2012})}\BibitemShut
  {NoStop}%
\bibitem [{\citenamefont {Aarts}\ \emph {et~al.}(2014)\citenamefont {Aarts},
  \citenamefont {Allton}, \citenamefont {Harris}, \citenamefont {Kim},
  \citenamefont {Lombardo}, \citenamefont {Ryan},\ and\ \citenamefont
  {Skullerud}}]{Aarts:2014cda}%
  \BibitemOpen
  \bibfield  {author} {\bibinfo {author} {\bibfnamefont {G.}~\bibnamefont
  {Aarts}}, \bibinfo {author} {\bibfnamefont {C.}~\bibnamefont {Allton}},
  \bibinfo {author} {\bibfnamefont {T.}~\bibnamefont {Harris}}, \bibinfo
  {author} {\bibfnamefont {S.}~\bibnamefont {Kim}}, \bibinfo {author}
  {\bibfnamefont {M.~P.}\ \bibnamefont {Lombardo}}, \bibinfo {author}
  {\bibfnamefont {S.~M.}\ \bibnamefont {Ryan}}, \ and\ \bibinfo {author}
  {\bibfnamefont {J.-I.}\ \bibnamefont {Skullerud}},\ }\href@noop {} {\bibfield
   {journal} {\bibinfo  {journal} {JHEP}\ }\textbf {\bibinfo {volume} {07}},\
  \bibinfo {pages} {097} (\bibinfo {year} {2014})}\BibitemShut {NoStop}%
\bibitem [{\citenamefont {Liu}\ and\ \citenamefont {Rapp}(2020)}]{Liu:2016ysz}%
  \BibitemOpen
  \bibfield  {author} {\bibinfo {author} {\bibfnamefont {S.~Y.~F.}\
  \bibnamefont {Liu}}\ and\ \bibinfo {author} {\bibfnamefont {R.}~\bibnamefont
  {Rapp}},\ }\href {\doibase 10.1140/epja/s10050-020-00024-z} {\bibfield
  {journal} {\bibinfo  {journal} {Eur. Phys. J. A}\ }\textbf {\bibinfo {volume}
  {56}},\ \bibinfo {pages} {44} (\bibinfo {year} {2020})},\ \Eprint
  {http://arxiv.org/abs/1612.09138} {arXiv:1612.09138 [nucl-th]} \BibitemShut
  {NoStop}%
\bibitem [{\citenamefont {Liu}\ and\ \citenamefont {Rapp}(2018)}]{Liu:2017qah}%
  \BibitemOpen
  \bibfield  {author} {\bibinfo {author} {\bibfnamefont {S.~Y.~F.}\
  \bibnamefont {Liu}}\ and\ \bibinfo {author} {\bibfnamefont {R.}~\bibnamefont
  {Rapp}},\ }\href@noop {} {\bibfield  {journal} {\bibinfo  {journal} {Phys.
  Rev. C}\ }\textbf {\bibinfo {volume} {97}},\ \bibinfo {pages} {034918}
  (\bibinfo {year} {2018})}\BibitemShut {NoStop}%
\bibitem [{\citenamefont {Rothkopf}(2020)}]{Rothkopf:2019ipj}%
  \BibitemOpen
  \bibfield  {author} {\bibinfo {author} {\bibfnamefont {A.}~\bibnamefont
  {Rothkopf}},\ }\href {\doibase 10.1016/j.physrep.2020.02.006} {\bibfield
  {journal} {\bibinfo  {journal} {Phys. Rept.}\ }\textbf {\bibinfo {volume}
  {858}},\ \bibinfo {pages} {1} (\bibinfo {year} {2020})},\ \Eprint
  {http://arxiv.org/abs/1912.02253} {arXiv:1912.02253 [hep-ph]} \BibitemShut
  {NoStop}%
\bibitem [{\citenamefont {Andronic}\ \emph {et~al.}(2024)\citenamefont
  {Andronic} \emph {et~al.}}]{Andronic:2024oxz}%
  \BibitemOpen
  \bibfield  {author} {\bibinfo {author} {\bibfnamefont {A.}~\bibnamefont
  {Andronic}} \emph {et~al.},\ }\href {\doibase
  10.1140/epja/s10050-024-01306-6} {\bibfield  {journal} {\bibinfo  {journal}
  {Eur. Phys. J. A}\ }\textbf {\bibinfo {volume} {60}},\ \bibinfo {pages} {88}
  (\bibinfo {year} {2024})},\ \Eprint {http://arxiv.org/abs/2402.04366}
  {arXiv:2402.04366 [nucl-th]} \BibitemShut {NoStop}%
\bibitem [{\citenamefont {Haglin}\ and\ \citenamefont
  {Gale}(1994)}]{Haglin:1994ap}%
  \BibitemOpen
  \bibfield  {author} {\bibinfo {author} {\bibfnamefont {K.~L.}\ \bibnamefont
  {Haglin}}\ and\ \bibinfo {author} {\bibfnamefont {C.}~\bibnamefont {Gale}},\
  }\href {\doibase 10.1016/0550-3213(94)90519-3} {\bibfield  {journal}
  {\bibinfo  {journal} {Nucl. Phys. B}\ }\textbf {\bibinfo {volume} {421}},\
  \bibinfo {pages} {613} (\bibinfo {year} {1994})},\ \Eprint
  {http://arxiv.org/abs/nucl-th/9401003} {arXiv:nucl-th/9401003} \BibitemShut
  {NoStop}%
\bibitem [{\citenamefont {Song}(1996)}]{Song:1996gw}%
  \BibitemOpen
  \bibfield  {author} {\bibinfo {author} {\bibfnamefont {C.}~\bibnamefont
  {Song}},\ }\href {\doibase 10.1016/0370-2693(96)01153-7} {\bibfield
  {journal} {\bibinfo  {journal} {Phys. Lett. B}\ }\textbf {\bibinfo {volume}
  {388}},\ \bibinfo {pages} {141} (\bibinfo {year} {1996})},\ \Eprint
  {http://arxiv.org/abs/hep-ph/9603259} {arXiv:hep-ph/9603259} \BibitemShut
  {NoStop}%
\bibitem [{\citenamefont {Vujanovic}\ \emph {et~al.}(2009)\citenamefont
  {Vujanovic}, \citenamefont {Ruppert},\ and\ \citenamefont
  {Gale}}]{Vujanovic:2009wr}%
  \BibitemOpen
  \bibfield  {author} {\bibinfo {author} {\bibfnamefont {G.}~\bibnamefont
  {Vujanovic}}, \bibinfo {author} {\bibfnamefont {J.}~\bibnamefont {Ruppert}},
  \ and\ \bibinfo {author} {\bibfnamefont {C.}~\bibnamefont {Gale}},\ }\href
  {\doibase 10.1103/PhysRevC.80.044907} {\bibfield  {journal} {\bibinfo
  {journal} {Phys. Rev. C}\ }\textbf {\bibinfo {volume} {80}},\ \bibinfo
  {pages} {044907} (\bibinfo {year} {2009})},\ \Eprint
  {http://arxiv.org/abs/0907.5385} {arXiv:0907.5385 [nucl-th]} \BibitemShut
  {NoStop}%
\bibitem [{\citenamefont {Li}\ and\ \citenamefont {Ko}(1995)}]{Li:1994cj}%
  \BibitemOpen
  \bibfield  {author} {\bibinfo {author} {\bibfnamefont {G.-Q.}\ \bibnamefont
  {Li}}\ and\ \bibinfo {author} {\bibfnamefont {C.~M.}\ \bibnamefont {Ko}},\
  }\href {\doibase 10.1016/0375-9474(94)00500-M} {\bibfield  {journal}
  {\bibinfo  {journal} {Nucl. Phys. A}\ }\textbf {\bibinfo {volume} {582}},\
  \bibinfo {pages} {731} (\bibinfo {year} {1995})},\ \Eprint
  {http://arxiv.org/abs/nucl-th/9407016} {arXiv:nucl-th/9407016} \BibitemShut
  {NoStop}%
\bibitem [{\citenamefont {Ko}\ \emph {et~al.}(1997)\citenamefont {Ko},
  \citenamefont {Koch},\ and\ \citenamefont {Li}}]{Ko:1997kb}%
  \BibitemOpen
  \bibfield  {author} {\bibinfo {author} {\bibfnamefont {C.~M.}\ \bibnamefont
  {Ko}}, \bibinfo {author} {\bibfnamefont {V.}~\bibnamefont {Koch}}, \ and\
  \bibinfo {author} {\bibfnamefont {G.-Q.}\ \bibnamefont {Li}},\ }\href
  {\doibase 10.1146/annurev.nucl.47.1.505} {\bibfield  {journal} {\bibinfo
  {journal} {Ann. Rev. Nucl. Part. Sci.}\ }\textbf {\bibinfo {volume} {47}},\
  \bibinfo {pages} {505} (\bibinfo {year} {1997})},\ \Eprint
  {http://arxiv.org/abs/nucl-th/9702016} {arXiv:nucl-th/9702016} \BibitemShut
  {NoStop}%
\bibitem [{\citenamefont {Pal}\ \emph {et~al.}(2002)\citenamefont {Pal},
  \citenamefont {Ko},\ and\ \citenamefont {Lin}}]{Pal:2002aw}%
  \BibitemOpen
  \bibfield  {author} {\bibinfo {author} {\bibfnamefont {S.}~\bibnamefont
  {Pal}}, \bibinfo {author} {\bibfnamefont {C.~M.}\ \bibnamefont {Ko}}, \ and\
  \bibinfo {author} {\bibfnamefont {Z.-w.}\ \bibnamefont {Lin}},\ }\href
  {\doibase 10.1016/S0375-9474(02)00992-2} {\bibfield  {journal} {\bibinfo
  {journal} {Nucl. Phys. A}\ }\textbf {\bibinfo {volume} {707}},\ \bibinfo
  {pages} {525} (\bibinfo {year} {2002})},\ \Eprint
  {http://arxiv.org/abs/nucl-th/0202086} {arXiv:nucl-th/0202086} \BibitemShut
  {NoStop}%
\bibitem [{\citenamefont {Coleman}\ \emph {et~al.}(1969)\citenamefont
  {Coleman}, \citenamefont {Wess},\ and\ \citenamefont
  {Zumino}}]{Coleman:1969sm}%
  \BibitemOpen
  \bibfield  {author} {\bibinfo {author} {\bibfnamefont {S.~R.}\ \bibnamefont
  {Coleman}}, \bibinfo {author} {\bibfnamefont {J.}~\bibnamefont {Wess}}, \
  and\ \bibinfo {author} {\bibfnamefont {B.}~\bibnamefont {Zumino}},\ }\href
  {\doibase 10.1103/PhysRev.177.2239} {\bibfield  {journal} {\bibinfo
  {journal} {Phys. Rev.}\ }\textbf {\bibinfo {volume} {177}},\ \bibinfo {pages}
  {2239} (\bibinfo {year} {1969})}\BibitemShut {NoStop}%
\bibitem [{\citenamefont {Callan}\ \emph {et~al.}(1969)\citenamefont {Callan},
  \citenamefont {Coleman}, \citenamefont {Wess},\ and\ \citenamefont
  {Zumino}}]{Callan:1969sn}%
  \BibitemOpen
  \bibfield  {author} {\bibinfo {author} {\bibfnamefont {C.~G.}\ \bibnamefont
  {Callan}, \bibfnamefont {Jr.}}, \bibinfo {author} {\bibfnamefont {S.~R.}\
  \bibnamefont {Coleman}}, \bibinfo {author} {\bibfnamefont {J.}~\bibnamefont
  {Wess}}, \ and\ \bibinfo {author} {\bibfnamefont {B.}~\bibnamefont
  {Zumino}},\ }\href {\doibase 10.1103/PhysRev.177.2247} {\bibfield  {journal}
  {\bibinfo  {journal} {Phys. Rev.}\ }\textbf {\bibinfo {volume} {177}},\
  \bibinfo {pages} {2247} (\bibinfo {year} {1969})}\BibitemShut {NoStop}%
\bibitem [{\citenamefont {Wess}\ and\ \citenamefont
  {Zumino}(1971)}]{Wess:1971yu}%
  \BibitemOpen
  \bibfield  {author} {\bibinfo {author} {\bibfnamefont {J.}~\bibnamefont
  {Wess}}\ and\ \bibinfo {author} {\bibfnamefont {B.}~\bibnamefont {Zumino}},\
  }\href {\doibase 10.1016/0370-2693(71)90582-X} {\bibfield  {journal}
  {\bibinfo  {journal} {Phys. Lett. B}\ }\textbf {\bibinfo {volume} {37}},\
  \bibinfo {pages} {95} (\bibinfo {year} {1971})}\BibitemShut {NoStop}%
\bibitem [{\citenamefont {Ecker}\ \emph
  {et~al.}(1989{\natexlab{a}})\citenamefont {Ecker}, \citenamefont {Gasser},
  \citenamefont {Pich},\ and\ \citenamefont {de~Rafael}}]{Ecker:1988te}%
  \BibitemOpen
  \bibfield  {author} {\bibinfo {author} {\bibfnamefont {G.}~\bibnamefont
  {Ecker}}, \bibinfo {author} {\bibfnamefont {J.}~\bibnamefont {Gasser}},
  \bibinfo {author} {\bibfnamefont {A.}~\bibnamefont {Pich}}, \ and\ \bibinfo
  {author} {\bibfnamefont {E.}~\bibnamefont {de~Rafael}},\ }\href {\doibase
  10.1016/0550-3213(89)90346-5} {\bibfield  {journal} {\bibinfo  {journal}
  {Nucl. Phys. B}\ }\textbf {\bibinfo {volume} {321}},\ \bibinfo {pages} {311}
  (\bibinfo {year} {1989}{\natexlab{a}})}\BibitemShut {NoStop}%
\bibitem [{\citenamefont {Ecker}\ \emph
  {et~al.}(1989{\natexlab{b}})\citenamefont {Ecker}, \citenamefont {Gasser},
  \citenamefont {Leutwyler}, \citenamefont {Pich},\ and\ \citenamefont
  {de~Rafael}}]{Ecker:1989yg}%
  \BibitemOpen
  \bibfield  {author} {\bibinfo {author} {\bibfnamefont {G.}~\bibnamefont
  {Ecker}}, \bibinfo {author} {\bibfnamefont {J.}~\bibnamefont {Gasser}},
  \bibinfo {author} {\bibfnamefont {H.}~\bibnamefont {Leutwyler}}, \bibinfo
  {author} {\bibfnamefont {A.}~\bibnamefont {Pich}}, \ and\ \bibinfo {author}
  {\bibfnamefont {E.}~\bibnamefont {de~Rafael}},\ }\href {\doibase
  10.1016/0370-2693(89)91627-4} {\bibfield  {journal} {\bibinfo  {journal}
  {Phys. Lett. B}\ }\textbf {\bibinfo {volume} {223}},\ \bibinfo {pages} {425}
  (\bibinfo {year} {1989}{\natexlab{b}})}\BibitemShut {NoStop}%
\bibitem [{\citenamefont {Meissner}(1988)}]{Meissner:1987ge}%
  \BibitemOpen
  \bibfield  {author} {\bibinfo {author} {\bibfnamefont {U.~G.}\ \bibnamefont
  {Meissner}},\ }\href {\doibase 10.1016/0370-1573(88)90090-7} {\bibfield
  {journal} {\bibinfo  {journal} {Phys. Rept.}\ }\textbf {\bibinfo {volume}
  {161}},\ \bibinfo {pages} {213} (\bibinfo {year} {1988})}\BibitemShut
  {NoStop}%
\bibitem [{\citenamefont {Bando}\ \emph {et~al.}(1988)\citenamefont {Bando},
  \citenamefont {Kugo},\ and\ \citenamefont {Yamawaki}}]{Bando:1987br}%
  \BibitemOpen
  \bibfield  {author} {\bibinfo {author} {\bibfnamefont {M.}~\bibnamefont
  {Bando}}, \bibinfo {author} {\bibfnamefont {T.}~\bibnamefont {Kugo}}, \ and\
  \bibinfo {author} {\bibfnamefont {K.}~\bibnamefont {Yamawaki}},\ }\href
  {\doibase 10.1016/0370-1573(88)90019-1} {\bibfield  {journal} {\bibinfo
  {journal} {Phys. Rept.}\ }\textbf {\bibinfo {volume} {164}},\ \bibinfo
  {pages} {217} (\bibinfo {year} {1988})}\BibitemShut {NoStop}%
\bibitem [{\citenamefont {Gale}\ and\ \citenamefont
  {Kapusta}(1991)}]{Gale:1990pn}%
  \BibitemOpen
  \bibfield  {author} {\bibinfo {author} {\bibfnamefont {C.}~\bibnamefont
  {Gale}}\ and\ \bibinfo {author} {\bibfnamefont {J.~I.}\ \bibnamefont
  {Kapusta}},\ }\href {\doibase 10.1016/0550-3213(91)90459-B} {\bibfield
  {journal} {\bibinfo  {journal} {Nucl. Phys. B}\ }\textbf {\bibinfo {volume}
  {357}},\ \bibinfo {pages} {65} (\bibinfo {year} {1991})}\BibitemShut
  {NoStop}%
\bibitem [{\citenamefont {Hohler}\ and\ \citenamefont
  {Rapp}(2014)}]{Hohler:2013ena}%
  \BibitemOpen
  \bibfield  {author} {\bibinfo {author} {\bibfnamefont {P.~M.}\ \bibnamefont
  {Hohler}}\ and\ \bibinfo {author} {\bibfnamefont {R.}~\bibnamefont {Rapp}},\
  }\href {\doibase 10.1103/PhysRevD.89.125013} {\bibfield  {journal} {\bibinfo
  {journal} {Phys. Rev. D}\ }\textbf {\bibinfo {volume} {89}},\ \bibinfo
  {pages} {125013} (\bibinfo {year} {2014})},\ \Eprint
  {http://arxiv.org/abs/1309.7036} {arXiv:1309.7036 [hep-ph]} \BibitemShut
  {NoStop}%
\bibitem [{\citenamefont {Hohler}\ and\ \citenamefont
  {Rapp}(2016)}]{Hohler:2015iba}%
  \BibitemOpen
  \bibfield  {author} {\bibinfo {author} {\bibfnamefont {P.~M.}\ \bibnamefont
  {Hohler}}\ and\ \bibinfo {author} {\bibfnamefont {R.}~\bibnamefont {Rapp}},\
  }\href {\doibase 10.1016/j.aop.2016.01.018} {\bibfield  {journal} {\bibinfo
  {journal} {Annals Phys.}\ }\textbf {\bibinfo {volume} {368}},\ \bibinfo
  {pages} {70} (\bibinfo {year} {2016})},\ \Eprint
  {http://arxiv.org/abs/1510.00454} {arXiv:1510.00454 [hep-ph]} \BibitemShut
  {NoStop}%
\bibitem [{\citenamefont {Cabrera}\ and\ \citenamefont
  {Vicente~Vacas}(2003)}]{Cabrera:2002hc}%
  \BibitemOpen
  \bibfield  {author} {\bibinfo {author} {\bibfnamefont {D.}~\bibnamefont
  {Cabrera}}\ and\ \bibinfo {author} {\bibfnamefont {M.~J.}\ \bibnamefont
  {Vicente~Vacas}},\ }\href {\doibase 10.1103/PhysRevC.67.045203} {\bibfield
  {journal} {\bibinfo  {journal} {Phys. Rev. C}\ }\textbf {\bibinfo {volume}
  {67}},\ \bibinfo {pages} {045203} (\bibinfo {year} {2003})},\ \Eprint
  {http://arxiv.org/abs/nucl-th/0205075} {arXiv:nucl-th/0205075} \BibitemShut
  {NoStop}%
\bibitem [{\citenamefont {Srednicki}(2007)}]{Srednicki:2007qs}%
  \BibitemOpen
  \bibfield  {author} {\bibinfo {author} {\bibfnamefont {M.}~\bibnamefont
  {Srednicki}},\ }\href@noop {} {\emph {\bibinfo {title} {{Quantum field
  theory}}}}\ (\bibinfo  {publisher} {Cambridge University Press},\ \bibinfo
  {year} {2007})\BibitemShut {NoStop}%
\bibitem [{\citenamefont {Zacchi}\ \emph {et~al.}(2015)\citenamefont {Zacchi},
  \citenamefont {Stiele},\ and\ \citenamefont
  {Schaffner-Bielich}}]{Zacchi:2015lwa}%
  \BibitemOpen
  \bibfield  {author} {\bibinfo {author} {\bibfnamefont {A.}~\bibnamefont
  {Zacchi}}, \bibinfo {author} {\bibfnamefont {R.}~\bibnamefont {Stiele}}, \
  and\ \bibinfo {author} {\bibfnamefont {J.}~\bibnamefont
  {Schaffner-Bielich}},\ }\href {\doibase 10.1103/PhysRevD.92.045022}
  {\bibfield  {journal} {\bibinfo  {journal} {Phys. Rev. D}\ }\textbf {\bibinfo
  {volume} {92}},\ \bibinfo {pages} {045022} (\bibinfo {year} {2015})},\
  \Eprint {http://arxiv.org/abs/1506.01868} {arXiv:1506.01868 [astro-ph.HE]}
  \BibitemShut {NoStop}%
\bibitem [{\citenamefont {Faessler}\ \emph {et~al.}(2004)\citenamefont
  {Faessler}, \citenamefont {Fuchs}, \citenamefont {Krivoruchenko},\ and\
  \citenamefont {Martemyanov}}]{Faessler:2002qb}%
  \BibitemOpen
  \bibfield  {author} {\bibinfo {author} {\bibfnamefont {A.}~\bibnamefont
  {Faessler}}, \bibinfo {author} {\bibfnamefont {C.}~\bibnamefont {Fuchs}},
  \bibinfo {author} {\bibfnamefont {M.~I.}\ \bibnamefont {Krivoruchenko}}, \
  and\ \bibinfo {author} {\bibfnamefont {B.~V.}\ \bibnamefont {Martemyanov}},\
  }\href {\doibase 10.1103/PhysRevLett.93.052301} {\bibfield  {journal}
  {\bibinfo  {journal} {Phys. Rev. Lett.}\ }\textbf {\bibinfo {volume} {93}},\
  \bibinfo {pages} {052301} (\bibinfo {year} {2004})},\ \Eprint
  {http://arxiv.org/abs/nucl-th/0212064} {arXiv:nucl-th/0212064} \BibitemShut
  {NoStop}%
\bibitem [{\citenamefont {Becattini}\ \emph
  {et~al.}(2019{\natexlab{b}})\citenamefont {Becattini}, \citenamefont {Cao},\
  and\ \citenamefont {Speranza}}]{Becattini:2019ntv}%
  \BibitemOpen
  \bibfield  {author} {\bibinfo {author} {\bibfnamefont {F.}~\bibnamefont
  {Becattini}}, \bibinfo {author} {\bibfnamefont {G.}~\bibnamefont {Cao}}, \
  and\ \bibinfo {author} {\bibfnamefont {E.}~\bibnamefont {Speranza}},\ }\href
  {\doibase 10.1140/epjc/s10052-019-7213-6} {\bibfield  {journal} {\bibinfo
  {journal} {Eur. Phys. J. C}\ }\textbf {\bibinfo {volume} {79}},\ \bibinfo
  {pages} {741} (\bibinfo {year} {2019}{\natexlab{b}})},\ \Eprint
  {http://arxiv.org/abs/1905.03123} {arXiv:1905.03123 [nucl-th]} \BibitemShut
  {NoStop}%
\bibitem [{\citenamefont {Shuryak}\ and\ \citenamefont
  {Zahed}(2004)}]{Shuryak:2004tx}%
  \BibitemOpen
  \bibfield  {author} {\bibinfo {author} {\bibfnamefont {E.~V.}\ \bibnamefont
  {Shuryak}}\ and\ \bibinfo {author} {\bibfnamefont {I.}~\bibnamefont
  {Zahed}},\ }\href@noop {} {\bibfield  {journal} {\bibinfo  {journal} {Phys.
  Rev. D}\ }\textbf {\bibinfo {volume} {70}},\ \bibinfo {pages} {054507}
  (\bibinfo {year} {2004})}\BibitemShut {NoStop}%
\bibitem [{\citenamefont {Pang}\ \emph {et~al.}(2016)\citenamefont {Pang},
  \citenamefont {Petersen}, \citenamefont {Wang},\ and\ \citenamefont
  {Wang}}]{Pang:2016igs}%
  \BibitemOpen
  \bibfield  {author} {\bibinfo {author} {\bibfnamefont {L.-G.}\ \bibnamefont
  {Pang}}, \bibinfo {author} {\bibfnamefont {H.}~\bibnamefont {Petersen}},
  \bibinfo {author} {\bibfnamefont {Q.}~\bibnamefont {Wang}}, \ and\ \bibinfo
  {author} {\bibfnamefont {X.-N.}\ \bibnamefont {Wang}},\ }\href@noop {}
  {\bibfield  {journal} {\bibinfo  {journal} {Phys. Rev. Lett.}\ }\textbf
  {\bibinfo {volume} {117}},\ \bibinfo {pages} {192301} (\bibinfo {year}
  {2016})}\BibitemShut {NoStop}%
\bibitem [{\citenamefont {Adamczyk}\ \emph
  {et~al.}(2017{\natexlab{b}})\citenamefont {Adamczyk} \emph
  {et~al.}}]{STAR:2017sal}%
  \BibitemOpen
  \bibfield  {author} {\bibinfo {author} {\bibfnamefont {L.}~\bibnamefont
  {Adamczyk}} \emph {et~al.} (\bibinfo {collaboration} {STAR}),\ }\href
  {\doibase 10.1103/PhysRevC.96.044904} {\bibfield  {journal} {\bibinfo
  {journal} {Phys. Rev. C}\ }\textbf {\bibinfo {volume} {96}},\ \bibinfo
  {pages} {044904} (\bibinfo {year} {2017}{\natexlab{b}})},\ \Eprint
  {http://arxiv.org/abs/1701.07065} {arXiv:1701.07065 [nucl-ex]} \BibitemShut
  {NoStop}%
\bibitem [{\citenamefont {Wu}\ \emph {et~al.}(2021)\citenamefont {Wu},
  \citenamefont {Du}, \citenamefont {Sibila},\ and\ \citenamefont
  {Rapp}}]{Wu:2020zbx}%
  \BibitemOpen
  \bibfield  {author} {\bibinfo {author} {\bibfnamefont {B.}~\bibnamefont
  {Wu}}, \bibinfo {author} {\bibfnamefont {X.}~\bibnamefont {Du}}, \bibinfo
  {author} {\bibfnamefont {M.}~\bibnamefont {Sibila}}, \ and\ \bibinfo {author}
  {\bibfnamefont {R.}~\bibnamefont {Rapp}},\ }\href {\doibase
  10.1140/epja/s10050-021-00623-4} {\bibfield  {journal} {\bibinfo  {journal}
  {Eur. Phys. J. A}\ }\textbf {\bibinfo {volume} {57}},\ \bibinfo {pages} {122}
  (\bibinfo {year} {2021})},\ \bibinfo {note} {[Erratum: Eur.Phys.J.A 57, 314
  (2021)]},\ \Eprint {http://arxiv.org/abs/2006.09945} {arXiv:2006.09945
  [nucl-th]} \BibitemShut {NoStop}%
\bibitem [{\citenamefont {Fu}\ \emph {et~al.}(2020)\citenamefont {Fu},
  \citenamefont {Pawlowski},\ and\ \citenamefont {Rennecke}}]{Fu:2019hdw}%
  \BibitemOpen
  \bibfield  {author} {\bibinfo {author} {\bibfnamefont {W.-j.}\ \bibnamefont
  {Fu}}, \bibinfo {author} {\bibfnamefont {J.~M.}\ \bibnamefont {Pawlowski}}, \
  and\ \bibinfo {author} {\bibfnamefont {F.}~\bibnamefont {Rennecke}},\ }\href
  {\doibase 10.1103/PhysRevD.101.054032} {\bibfield  {journal} {\bibinfo
  {journal} {Phys. Rev. D}\ }\textbf {\bibinfo {volume} {101}},\ \bibinfo
  {pages} {054032} (\bibinfo {year} {2020})},\ \Eprint
  {http://arxiv.org/abs/1909.02991} {arXiv:1909.02991 [hep-ph]} \BibitemShut
  {NoStop}%
\end{thebibliography}%
\pagebreak

\end{document}